\documentclass[twocolumn]{aastex631}
\usepackage{comment}
\usepackage{gensymb}
\usepackage{amsmath}
\usepackage{{booktabs}}
\usepackage{savesym}
\savesymbol{tablenum}
\usepackage{siunitx}
\restoresymbol{SIX}{tablenum}
\usepackage{siunitx}
\usepackage[mathlines]{lineno}

\graphicspath{{./}{figures/}}

\begin{document}
 
\title{Dwarf Galaxies with Radio-excess AGNs in the VLA Sky Survey}
\author{John-Michael Eberhard}
\affiliation{eXtreme Gravity Institute, Department of Physics, Montana State University, Bozeman, MT 59717, USA}
\author[0000-0001-7158-614X]{Amy E. Reines}
\affiliation{eXtreme Gravity Institute, Department of Physics, Montana State University, Bozeman, MT 59717, USA}
\author[0000-0003-1436-7658]{Hansung B. Gim}
\affiliation{eXtreme Gravity Institute, Department of Physics, Montana State University, Bozeman, MT 59717, USA}
\author[0000-0003-2511-2060]{Jeremy Darling}
\affiliation{Center for Astrophysics and Space Astronomy, Department of Astrophysical and Planetary Sciences, University of Colorado, 389 UCB, Boulder, CO 80309-0389, USA}
\author[0000-0002-5612-3427]{Jenny E. Greene}
\affiliation{Department of Astrophysical Sciences, Princeton University, Princeton, NJ
08544, USA}

\begin{abstract}
We present a systematic search for radio active galactic nuclei (AGNs) in dwarf galaxies using recent observations taken by the Very Large Array Sky Survey (VLASS). To select these objects, we first establish a criterion to identify radio-excess AGNs using the infrared-radio correlation (IRRC) parameter, $q$, that describes the tight relation between radio and IR emission in star forming (SF) galaxies. We find a $2\sigma$ threshold of $q < 1.94$ to select radio-excess AGNs, which is derived from a sample of $\sim 7,000$ galaxies across the full mass range in the NASA-Sloan Atlas (NSA) that have radio and IR detections from VLASS and the {\it Wide-Field Infrared Survey Explorer}, respectively. We create catalogs of radio-excess AGNs and SF galaxies and make these available to the community. Applying our criterion to dwarf galaxies with stellar masses $M_\star \lesssim 3 \times 10^9 M_\odot$ and redshifts $z \le 0.15$, and carefully removing interlopers, we find 10 radio-excess AGNs with radio-optical positional offsets between $\sim 0$ and $2\farcs3$ (0 - 2.7 kpc). Based on statistical arguments and emission line diagnostics, we expect the majority of these radio-excess AGNs to be associated with the dwarf host galaxies rather than background AGNs. Five of the objects have evidence for hosting AGNs at other wavelengths, and 5 objects are identified as AGNs in dwarf galaxies for the first time. {We also identify {8} variable radio sources in dwarf galaxies by comparing the VLASS {epoch 1 and epoch 2} 
observations to FIRST detections presented in \cite{reines2020}.}

\end{abstract}



\section{Introduction} \label{sec:intro}





The majority of black holes (BHs) belong to one of two populations: stellar mass BHs with masses $\sim10\:M_\odot$ and supermassive BHs (SMBHs) with masses $\sim10^{6-9}M_\odot$. BHs with masses between these extremes are often referred to as intermediate mass BHs (IMBHs) and while hundreds of IMBHs with masses between $\sim10^{5-6}M_\odot$ have been discovered in dwarf galaxies (e.g., \citealt{reines2022}), evidence for BHs with masses $\sim10^{2-5}M_\odot$ is scarce (\citealt{greene2020} and references therein). 

In addition to their rarity, IMBHs are objects of interest because they are critical to better understanding the origin of SMBHs. For SMBHs to reach their immense sizes, it is widely accepted that they evolved from smaller BHs, known as seeds, that grew through accretion and mergers. Different seeding models have been proposed (e.g., \citealt{vol2008}; \citealt{vol2009}; \citealt{agarwal2012}; \citealt{ricarte2018}; \citealt{rose2022}), but it remains unclear as to which models are the most accurate, or if a combination of models is closest to reality. Each seeding model predicts a different size and quantity of seed BHs in the early universe, and consequently, predicts a different distribution of modern-day IMBHs. Therefore, the discovery of more IMBHs may aid in constraining seeding models and improve our understanding of the way that SMBHs form and evolve.

We study dwarf galaxies to optimize the likelihood of discovering IMBHs, since observational scaling relationships have shown that less massive galaxies host central BHs of lesser mass (e.g., \citealt{kor2013}; \citealt{mcconnell2013}; \citealt{reines2015}; \citealt{bentz2018}; \citealt{schutte2019}). According to these relations, the masses of the BHs near the center of dwarf galaxies lie within the IMBH mass range.

AGNs in dwarf galaxies can be studied over a variety of wavelengths, and each wavelength regime has proven to have a unique set of advantages and disadvantages. In the optical regime, AGNs have been identified using narrow emission line diagnostics \citep{reines2013,moran2014, molina2021,salehirad2022} and by searching for broad H$\alpha$ emission (e.g., \citealt{greene2007}; \citealt{dong2012}; \citealt{reines2013}). The presence of broad H$\alpha$ in conjunction with AGN-like narrow line ratios has been shown to be  effective in discerning between AGNs and supernovae (SNe), which can also produce broad H$\alpha$ lines in dwarf galaxies (\citealt{baldassare2016}).
One limitation of optical diagnostics is that they typically can only find BHs that are radiating at large fractions of their Eddington luminosity, which is fairly uncommon. 
Moreover, dust can extinguish optical light emitted from AGNs. 
X-ray emission is beneficial in detecting low-luminosity AGNs (e.g., \citealt{lemons2015}; \citealt{reines2016}; \citealt{mezcua2018}; \citealt{latimer2019}), 
but X-ray observations require large exposure times, which is not practical for large-scale searches. Mid-IR observations also have the potential to identify AGNs, but when studying dwarf galaxies, it has proven difficult to separate the emission from intense star formation from the emission due to AGNs (\citealt{hainline2016}; 
\citealt{satyapal2018}; \citealt{latimer2021b}). 

Radio searches for AGNs have some advantages compared to other wavelength regimes. AGNs almost always create radio emission at centimeter wavelengths (\citealt{ho2008} and references therein). {Strong radio signals are primarily produced by synchrotron emission in the form of radio jets. Weaker radio signals from AGNs can come from a variety of sources, including AGN driven winds, thermal free-free emission from photoionized gas, and coronal activity from the accretion disc (see \citealt{panessa2019}, and references therein)}.
Another advantage of using radio searches is that radio emission is nearly unaffected by interstellar dust extinction.
{These factors, combined with the substantial area covered by existing radio surveys, make the use of radio observations an excellent method for conducting large-scale searches for AGNs. For these reasons, previous radio searches have been used to identify AGNs in dwarf galaxies, using data collected by the Very Large Array \citep{mezcua2019, reines2020}}. 

One difficulty with working in the radio regime, however, is that radio emission is not unique to AGNs. 
Star formation (SF) creates radio emission that can mimic or mask the radio signals created by AGNs (e.g., \citealt{kimball2011}; \citealt{condon2013}; \citealt{zakamska2015}). SF creates HII regions, which in turn produce thermal emission via free-free emission (e.g., \citealt{caplan1986}), while SNRs and SNe create non-thermal radio emission via synchrotron radiation, as cosmic ray electrons from SNRs and SNe are accelerated in galactic magnetic fields (see the \citealt{condon1992} review).

In this work, we identify galaxies with AGNs by finding sources with radio fluxes from the VLA Sky Survey (VLASS) that lie significantly above the well-known IR-radio relation for star forming galaxies. {We derive a criterion for discerning between SF and AGNs using galaxies of all masses in the NASA-Sloan Atlas with detections in VLASS and WISE, and then apply our method to dwarf galaxies specifically.}
The paper is organized as follows: in \S\ref{sec:Data} we describe the data and catalogs we use in this work. In \S\ref{sec:radio-excess_AGNs_VLASS}, we explain how we match all the galaxies from the NASA-Sloan Atlas to VLASS radio sources and develop a cutoff between emission from SF and emission from radio-excess AGNs. We analyze the properties of all the radio-excess AGN hosts and compare them to results from the literature in \S\ref{sec:Properties_NSA-VLASS-WISE}. {In \S\ref{sec:Dwarfs}, we apply the cutoff found in \S\ref{sec:radio-excess_AGNs_VLASS} to lower mass galaxies to find dwarf galaxies with radio-excess AGNs. We analyze their properties and compare them to the full sample of radio-excess AGNs.} We conclude with a summary and a discussion of our discoveries in \S\ref{sec:Conclusions}. For consistency with \cite{reines2020}, we use $H_{0} = 73 \mathrm{\;km \; s^{-1}\;  Mpc^{-1}}$ throughout.

\section{Data}
\label{sec:Data}
In this paper, we match the radio sources detected by VLASS to galaxies in the NASA-Sloan Atlas (NSA), similar to how \cite{reines2020} matched radio sources from the Faint Images of the Radio Sky at Twenty-centimeters (FIRST) Survey to the NSA. However, several key differences between the data used here and \cite{reines2020} make this study worthwhile and novel. 

One major difference is the quality of VLASS compared to FIRST. FIRST had an angular resolution of 5\arcsec\ and a median RMS sensitivity of $\sim$100 $\mu$Jy/beam, and covered 10,000 square degrees of sky \citep{white1997}. 
VLASS, in contrast, has an angular resolution of 2\farcs5 and covers 33,885 square degrees of sky (all of the sky above $-40\degree$ declination). The RMS sensitivity per epoch of VLASS is comparable to FIRST (120 $\mu$Jy/beam), but the expected $1\sigma$ sensitivity is 70 $\mu$Jy/beam once its three planned epochs are complete \citep{lacy2020}. Furthermore, FIRST concluded its observations in 2011 and VLASS is currently still observing. The difference in time between observations also results in finding transient sources that were not bright enough for detection a decade ago.

{Due to its higher angular resolution, VLASS can separate single sources in FIRST into multiple sources and find radio lobes as radio sources that do not coincide with their host galaxies. We minimize the likelihood of matching radio sources to unrelated galaxies by utilizing a small cross-matching radius between surveys (see \S\ref{subsec:NSA-VLASS-WISE}).}

Another critical difference between this paper and \cite{reines2020} is the use of an updated NSA catalog. \cite{reines2020} utilized v0\textunderscore1\textunderscore2 of the NSA, which extended to a redshift of $z=0.055$, whereas the the version used in this paper (v1\textunderscore0\textunderscore1) extends to a redshift of $z=0.15$.
Between the more recent and higher quality radio data from VLASS and the larger sample of galaxies from the newer version of the NSA, we find new candidates for radio-excess AGNs within dwarf galaxies than those found by \cite{reines2020}.

\subsection{Very Large Array Sky Survey (VLASS)}
\label{subsec:VLASS}
VLASS is performed by the NSF’s Karl G. Jansky Very Large Array (VLA) in the B-configuration in the S-band (2-4 GHz). Observations began in September 2017 and the survey is expected to complete three epochs by the end of 2024. 
As of the time of this paper's submission, two of the three planned epochs are complete and data products have been provided as part of the Canadian Initiative for Radio Astronomy Data Analysis (CIRADA) program\footnote{CIRADA is funded by a grant from the Canada Foundation for Innovation 2017 Innovation Fund (Project 35999), as well as by the Provinces of Ontario, British Columbia, Alberta, Manitoba and Quebec.}, which partnered with National Radio Astronomy Observatory (NRAO)\footnote{The National Radio Astronomy Observatory is a facility of the National Science Foundation operated under cooperative agreement by Associated Universities, Inc.} and the Canadian Astronomy Data Centre (CADC).

We retrieve the entire epoch 2 catalog from the CIRADA web page and process the data to remove duplicate radio sources and images with poor quality flags. {Based on the beam size and positional uncertainty of VLASS images, the CIRADA catalog flags duplicates as sources within 2\arcsec  of another source. When duplicates are found, preference is given to the source with the highest ratio peak flux density to local rms. Restricting by the duplicate flag in the CIRADA catalog restricts the catalog to only include sources that are unique or preferred components.
By restricting with the quality flag, we remove every radio source whose peak brightness is less than 5 times larger than the local RMS of its respective image. Of the 2,995,271 sources in the original catalog, 1,125,370 (38\%) of the sources are removed due to their low quality flags and a further 174,749 (6\%) of the sources are removed based on their duplicate flags.} This leaves 1,695,152 radio sources in our sample with S/N cuts at the $5\sigma$ level.

\subsection{NASA-Sloan Atlas (NSA)}
\label{subsec:NSA}
The NASA-Sloan Atlas (NSA, v1\textunderscore0\textunderscore1) is a catalog of images and parameters for local galaxies, with redshift $z \le 0.15$, including data taken from surveys in the near-IR, optical, and UV regimes. The emphasis of the catalog is on the inclusion of parameters derived from the Sloan Digital Sky Survey (SDSS) and the Galaxy Evolution Explorer (GALEX, \citealt{york2000}; \citealt{blanton2005}). {The version of the NSA catalog we use is based on the observations in the SDSS DR11, with a sky coverage area of 14,555 square degrees.}  

Photometric measurements in the NSA are derived by making image mosaics and rephotometering the images from the $ugriz$ bands in SDSS and the far ultraviolet (FUV) and near ultraviolet (NUV) bands from GALEX.  The version used in this work utilizes elliptical Petrosian apertures to report the K-corrected absolute AB magnitudes in each of the bands. K-corrections are determined using the \textit{kcorrect} package \citep{blanton2007} {which fits SDSS and broadband Galaxy Evolution Explorer (GALEX) fluxes using templates based on the stellar population synthesis models of \cite{bruzual2003} and the nebular emission line models of \cite{kewley2001}. The stellar mass estimates that correspond to the best fits are included in the NSA.}  


The majority of redshifts in the NSA are derived from SDSS 
spectra, but in some cases redshifts are taken from other sources, including the NASA Extragalactic Database (NED), the Six-degree Field Galaxy Redshift Survey (sixdf), the Two-degree Field Galaxy Redshift Survey (twodf), the CfA Redshift Survey (ZCAT), and Arecibo Legacy Fast Arecibo L-Band Feed Array (ALFALFA).


\subsection{Wide-Field Infrared Survey Explorer (WISE)}
\label{subsec:WISE}
WISE is a NASA IR space telescope that performed an all-sky astronomical survey starting {January 2010}. WISE used a 40 cm diameter telescope in four IR bands: W1, W2, W3, and W4, which are centered at wavelengths of 3.4, 4.6, 12, and 22 $\mu$m, with angular resolutions of  6\farcs1, 6\farcs4, 6\farcs5, and 12\farcs0, respectively \citep{wright2010}.
Observations and S/N for each of the IR bands are recorded in the AllWISE source catalog\footnote{This publication makes use of data products from the Wide-field Infrared Survey Explorer, which is a joint project of the University of California, Los Angeles, and the Jet Propulsion Laboratory/California Institute of Technology, funded by the National Aeronautics and Space Administration.} \citep{allwisecat}, and are only included in the catalog if at least one of the four bands has S/N $>$ 5. We utilize the the ``profile-fitting'' photometric magnitudes from the AllWISE catalogs, which are optimized for unresolved sources.

\section{Finding Radio-excess AGN\lowercase{s} in VLASS}\label{sec:radio-excess_AGNs_VLASS}
{To distinguish between radio emission dominated by SF and emission dominated by AGNs, we compare the radio luminosities from VLASS to IR luminosities from WISE. We utilize the IR-radio correlation (IRRC) parameter $q$, as has been done in previous papers (e.g., \citealt{yun2001}; \citealt{condon2002}; \citealt{delmoro2013}; \citealt{delhaize2017}; \citealt{delv2021}) to find galaxies with radio emission significantly higher than expected from SF. 
We study the full sample of galaxies from the NSA in this section to better quantify the scatter and develop an accurate method for discerning between SF-related emission and emission from AGNs.}


\subsection{NSA-VLASS-WISE Sample}
\label{subsec:NSA-VLASS-WISE}
We retrieve the full catalog of galaxies in the NSA 
to create a parent sample of 641,409 galaxies. We cross-match the NSA and VLASS coordinates with a cross-matching radius of 2\farcs5, resulting in finding 13,688 galaxies within the NSA with associated VLASS detections. 

We then cross-match our sample of NSA galaxies with VLASS detections to the ALLWISE catalog, again employing a cross-matching radius of 2\farcs5 and selecting the closest IR source if multiple sources are found within the radius. 
{We require the WISE detections to be high-quality measurements with S/N $\ge$ 5 in the W4 band. This ensures an accurate distribution of IRRC values for the galaxies in our sample and enables us to determine a reliable cutoff for identifying radio-excess AGNs (see Section 3.3). We note that while we do not use the galaxies with S/N $<$ 5 to calculate the IRRC cutoff, we do include them in our AGN catalog provided in the appendix.}




{For the purposes of determining the IRRC distribution, we also remove sources that have underestimated fluxes in WISE. The ``profile-fitting" magnitudes we use from the WISE catalog are optimized for point-like sources, meaning that the fluxes of resolved sources are systematically underestimated. We therefore remove all extended sources from our sample, where we consider extended sources to be any sources with a W4 profile-fit photometry goodness-of-fit ($\chi^2$) above 3 (the same criterion used by the AllWISE catalog to flag extended sources). By applying this criterion, we remove 219 extended IR sources from our sample.}

{Once these sources are removed, we are left with a total of 6,772 galaxies with reliable W4 measurements. These galaxies have high values of S/N in all of the WISE bands. All have W2 and W3 detections with S/N $>$ 5. Only 6 galaxies have W1 magnitudes with S/N $<$ 5, and even these have S/N $\ge$ 4. Restricting by W4 therefore results in finding galaxies that have strong detections in all bands of WISE.}
We refer to the sample of galaxies in the NSA with VLASS and strong WISE detections as the NSA-VLASS-WISE sample.

\subsection{Removal of Mid-IR AGNs}
\label{subsec:remove_midIR}
\begin{figure}
 \includegraphics[width=\columnwidth]{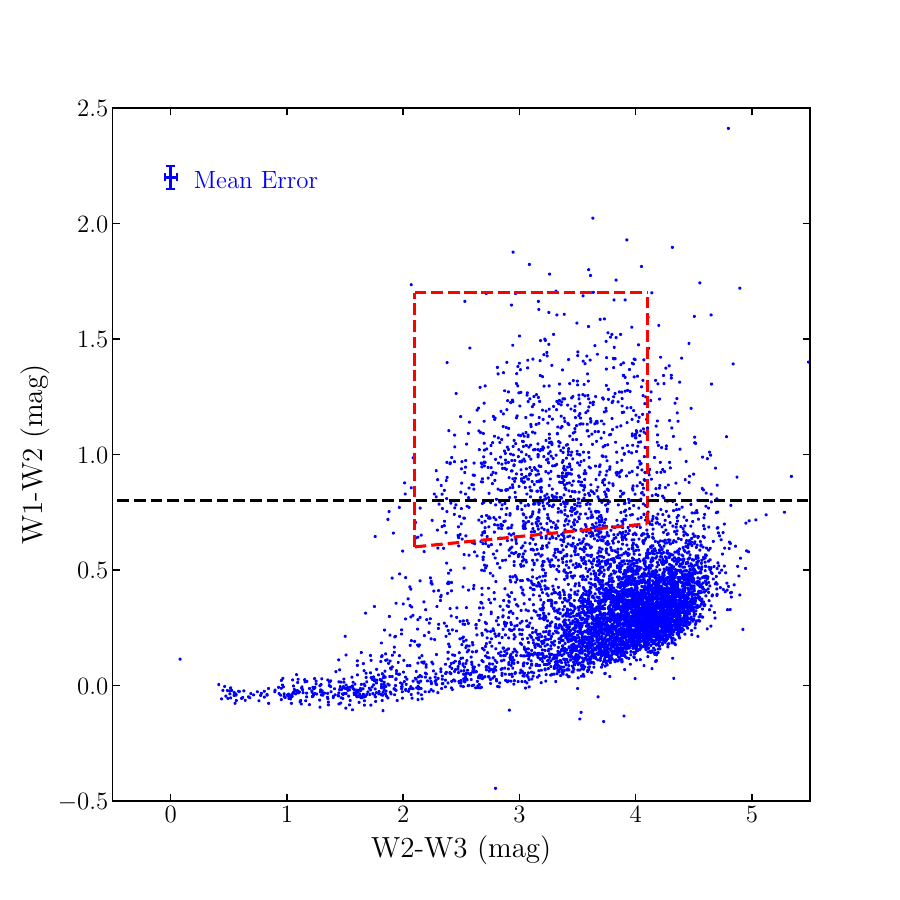}
 \caption{WISE color–color diagram for all galaxies in the NSA-VLASS-WISE sample. The \cite{jarrett2011} AGN selection box is shown in red and the \cite{stern2012} AGN cutoff line is shown in black. We remove all galaxies found within the selection box or the cutoff line to limit the effect of mid-IR AGN emission on the IR luminosities {used in the IRRC (\S \ref{subsec:IRRC})}. {The mean error bars for the data are shown in the upper left corner.}}
\label{fig:wise_color}
\end{figure}

To find radio-excess AGNs within our sample, we will need to differentiate between the emission from SF and the emission from radio-excess AGNs. This will be done by comparing the ratio of IR and radio luminosities (see \S\ref{subsec:IRRC}), under the assumption that the IR emission comes primarily from SF. To be able to make this assumption, we remove galaxies whose mid-IR emission is characteristic of AGNs. We compare the WISE IR colors (W1-W2 and W2-W3) 
for the galaxies in the NSA-VLASS-WISE sample in a color-color diagram (see Figure \ref{fig:wise_color}). We exclude mid-IR AGNs by removing galaxies that lie within the \cite{jarrett2011} AGN selection box or above the \cite{stern2012} AGN cutoff, finding {1,103} that fit this criteria.

{Given the high S/N for the W4 observations in our sample, we consider using the WISE four-band AGN selection criteria from \cite{mateos2012} instead of the criteria from \cite{jarrett2011} and \cite{stern2012}. However, the four-band method only identifies 620 objects mid-IR AGNs, compared to the 1,103 mid-IR AGNs found above. To minimize the likelihood of including mid-IR AGNs in our sample, we opt to use the combined \cite{jarrett2011} and \cite{stern2012} criteria, as this identifies a larger number of AGNs.}

{We note that the percentage of our sample (Section \ref{subsec:NSA-VLASS-WISE}) identified as mid-IR AGNs (16\%) is larger than is expected for the general galaxy population. This is likely due to our restriction on the W4 S/N, which means our sample preferentially contains galaxies with higher overall IR luminosities. Galaxies with higher IR luminosities in turn are more likely to host IR-selected AGNs, as evidenced by the sample studied in \cite{goulding2009}, which found 27\% of IR-bright galaxies to have AGNs.}


We also note that the \cite{jarrett2011} and \cite{stern2012} cutoffs were not established for radio-selected galaxies. So, while the cutoffs we use identify the majority of mid-IR AGNs in our sample, they may not properly account for the full scatter (both real and due to low S/N) of our radio-selected sample, and fail to find all of the mid-IR AGNs (e.g., \citealt{truebenbach2017}). Nevertheless, we work with the {5,669} remaining galaxies under the assumption that the IR emission for the galaxies is dominated by SF. 


\subsection{The IR-Radio Correlation (IRRC)}
\label{subsec:IRRC}
We develop a method for distinguishing between radio emission from SF and AGNs by analyzing the IRRC parameter for our sample.
The IRRC parameter $q$ describes the tight relation between radio and IR emission in star forming galaxies (SFGs) and takes the following form:
\begin{equation}
    q = \mathrm{log}\left[ \frac{\mathrm{L_{IR}}}{3.75\times10^{12} \; \mathrm{L_{1.4\;GHz}}}\right]
    \label{eq:IRRC}
\end{equation}
where $\mathrm{L_{IR}}$ is in units of W and  $\mathrm{L_{1.4\;GHz}}$ is in units of W Hz$^{-1}$ (e.g., \citealt{deJong1985}; \citealt{helou1985}). $\mathrm{L_{IR}}$ is frequently reported in terms of far-IR (FIR) wavelengths (40-120$\mathrm{\mu}$m) or total-IR (TIR) wavelengths (8-1000$\mathrm{\mu}$m).
For SFGs in the local universe, \cite{yun2001} found that $q_{FIR} = 2.34\pm 0.26$ dex for far-IR (FIR) wavelengths, and \cite{bell2003} found that $q_{TIR} = 2.64\pm 0.26$ dex for TIR wavelengths.

The IRRC parameter has also been used to find radio-excess AGNs by looking for sources that have excessively low values of $q$, indicating high amounts of radio emission that cannot be described by SF. However, the level of radio emission that is considered to be significantly high varies between studies.
\cite{condon2002} defined radio-excess AGNs are objects with radio emission 3 times larger than expected from the IRRC, while \cite{yun2001} defined radio-excess AGNs as objects with radio luminosities that are 5 times greater than the value predicted by the IRRC or larger. 
{The cutoff in \cite{yun2001} came from finding sources with $q$ values that were lower than 3 times the RMS scatter below the mean value of $q$ for the SFGs.}
\cite{bonzini2013} found an AGN cutoff using the full distribution of $q$ in their sample, defining radio-excess AGNs as objects with $q$ values that were lower than 2 standard deviations ($\sigma_{q}$) below than the mean value ($\mu_{q}$).
\cite{delmoro2013} and \cite{delhaize2017} performed a similar analysis, but used a cutoff of 3$\sigma_{q}$ below $\mu_{q}$.


\cite{delv2021} elaborated on the reason that such a variety of methods exist, explaining that defining the cutoff between AGN and SF radio emission is a difficult and somewhat arbitrary task. Changing the cutoff to a lower $q$ increases the purity of finding AGNs, but at the cost of finding all of the AGNs. \cite{delv2021} determined an ideal cutoff for $q$ by testing various thresholds and identifying the cutoffs that resulted in the best compromise between reaching low levels of cross-contamination between the AGN and SF populations and finding a large number of SFGs beneath the threshold. They found the cutoff at $q=q_{peak}-2\sigma$ to give the best trade-off between completeness and cross-contamination, where $q_{peak}$ and $\sigma$ are the peak value and standard deviation of the q distribution for the population, respectively.
We similarly adopt the $q=q_{peak}-2\sigma$ criteria in this work to find the maximum number of AGNs with minimal contamination from the SF population.

To calculate the distribution of $q$ values for our sample (using Equation \ref{eq:IRRC}) we convert the 3 GHz flux densities from VLASS to 1.4 GHz luminosities.
Just as \cite{novak2017} converted from VLA-COSMOS 3 GHz data to 1.4 data, we convert the 3 GHz flux densities from VLASS to 1.4 GHz luminosities with the following equation: 
\begin{equation}
    L_{\nu_{1}}= \frac{4\pi D_{L}^{2}}{(1 + z)^{1+\alpha}} \left(\frac{\nu_1}{\nu_2}\right)^{\alpha} S_{\nu_{2}}
    \label{eq:L_from_S}
\end{equation}

Here, $L_{\nu_{1}}$ is the rest-frame radio luminosity at {the rest-frame} frequency $\nu_1$, derived from the observed flux $S_{\nu_{2}}$ at the {observed-frame frequency} $\nu_2$, redshift $z$, luminosity distance $D_{L}$, and spectral index $\alpha$. To convert VLASS observations to radio luminosities that can be used in Equation \ref{eq:IRRC}, we set $\nu_{1}=1.4$ GHz and $\nu_{2}=3$ GHz.
As was done in \cite{novak2017}, we assume that each radio source in our sample has a radio spectrum that can be described as a simple power law ($S_{\nu}\propto \nu^{\alpha}$) where $S_{\nu}$ is the monochromatic flux at frequency $\nu$ and $\alpha$ is the spectral index, resulting in the K-correction of $K(z) = (1+z)^{-(1+\alpha)}$ seen in Equation \ref{eq:L_from_S}. 

\begin{figure}
\includegraphics[width=\columnwidth]{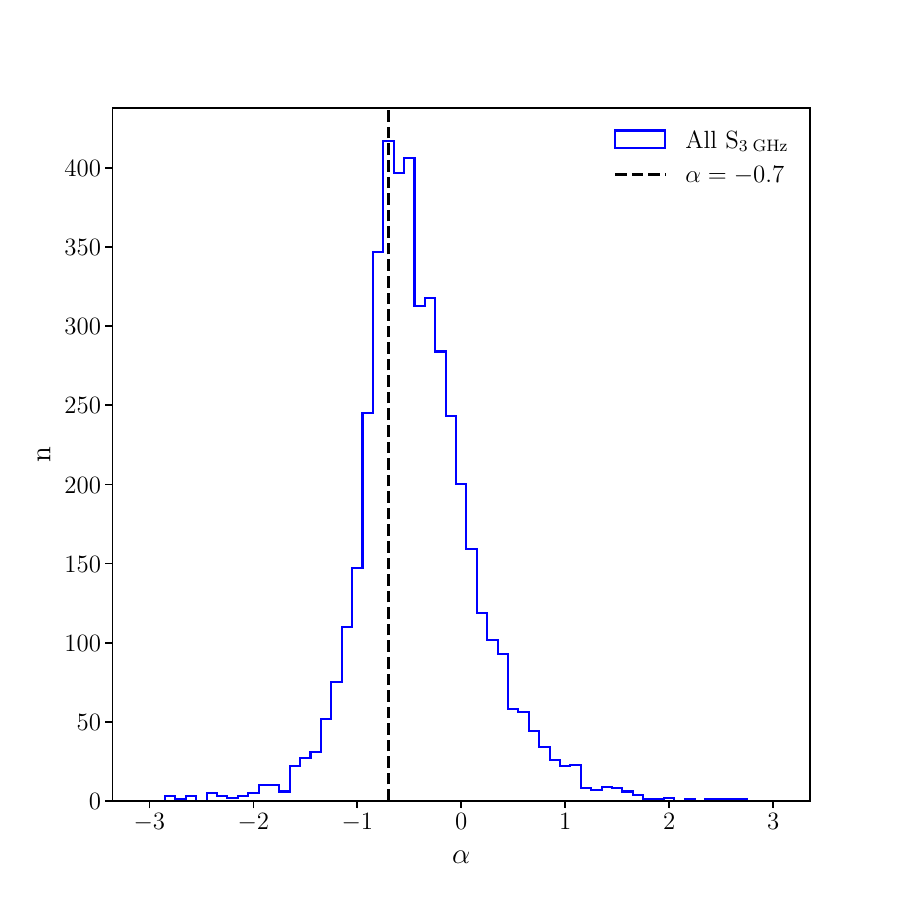}
 \caption{Distribution of spectral indices for the galaxies in the NSA-VLASS-WISE sample with corresponding FIRST observations. The distribution peaks at a similar value to the standard value of $\alpha=-0.7$, which is commonly assumed to be the median value for SFGs.
}
\label{fig:all_alpha}
\end{figure}

For $\sim$75\% of the sources,
we find $\alpha$ directly by matching the VLASS catalog to the FIRST catalog, with a cross-matching radius of 5\arcsec (the resolution of FIRST). The distribution of $\alpha$ for galaxies in the NSA-VLASS-WISE sample with FIRST detections is shown in Figure \ref{fig:all_alpha}. 
The values of the spectral indices are largely found between -2 and 2 and the distribution peaks at $\alpha \sim -0.7$. This peak value matches well with the standard $\alpha=-0.7$ that is commonly assumed in the literature (e.g., \citealt{condon2012}; \citealt{novak2017}; \citealt{smolcic2017b}).
As such, for the sources without FIRST detections, we assume a spectral index of $\alpha = -0.7$ in Equation \ref{eq:L_from_S}.



We estimate $\mathrm{L_{TIR}}$ using data from the W4 band of WISE, centered at a wavelength of 22 $\mu m$. \cite{hao2011} reported that $\mathrm{L_{TIR}} \approx 8.33 \: \mathrm{L_{25 \mu m}}$  and \cite{jarrett2013} showed that the flux densities at 22 $\mu m$ and 25 $\mu m$ are approximately equal. 
{These relations were established using samples of nearby galaxies that covered a wide range of color, luminosity, and stellar mass (including dwarf galaxies). Therefore, we deem it appropriate to use these relations on our full sample to convert directly from W4 luminosities to TIR luminosities.}



\begin{figure}
\includegraphics[width=\columnwidth]{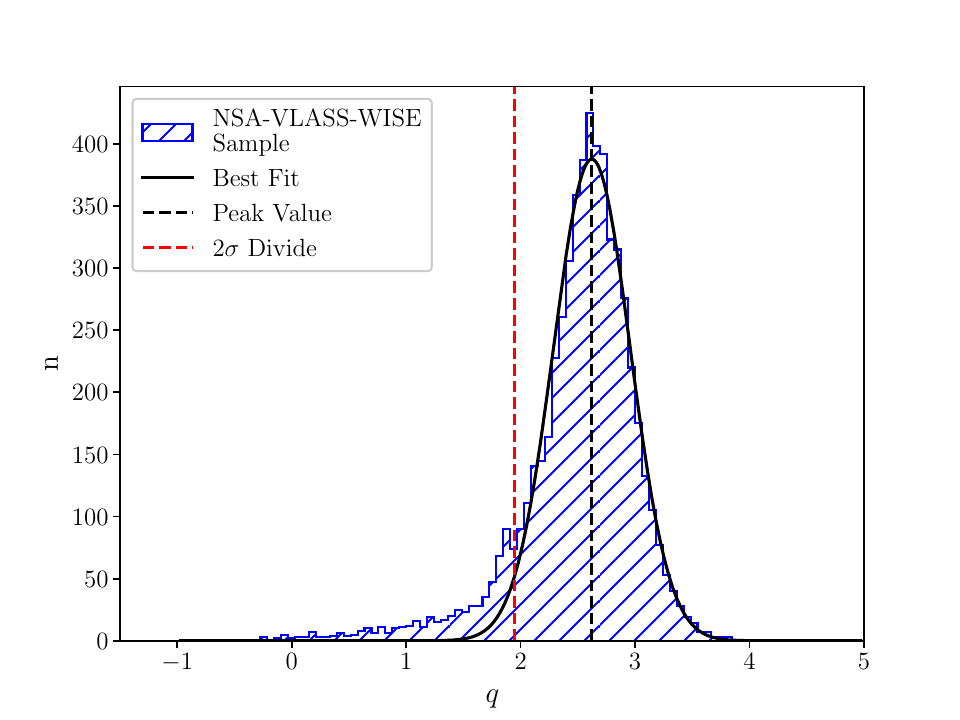}
 \caption{Distribution of $q$ for the galaxies in the NSA-VLASS-WISE sample that are not classified as mid-IR AGNs (see Figure \ref{fig:wise_color}). 
 {When a Gaussian is fit to the distribution, it peaks} at $q_{peak}=2.62$ and has a scatter of $\sigma=0.34$, resulting in a 2$\sigma$ threshold of $q < 1.94$ for identifying radio-excess AGNs.}
\label{fig:qdivide}
\end{figure}

Using the assumptions above, we calculate $q$ for all of the galaxies in the NSA-VLASS-WISE sample that are not classified as mid-IR AGNs and find the distribution shown in Figure \ref{fig:qdivide}. We fit a Gaussian to the distribution and find that the sample peaks at $q_{peak}=2.62$, matching well with the results from \cite{bell2003}. We find the scatter of the distribution to be $\sigma=0.34$, resulting in the 2$\sigma$ cutoff of $q < 1.94$. {When we apply this cutoff to our NSA-VLASS-WISE sample (excluding mid-IR AGNs), we find 625 (11\%) of the galaxies to host radio-excess AGNs}. 
{The percentage of radio AGNs found in our sample is larger than expected for the general galaxy population. This increase can be attributed to our selection of galaxies with VLASS detections with high S/N (see \S2.1). When we require the galaxies in our sample to have high quality radio data, we preferentially find objects that are louder in the radio regime, which results in finding a larger percentage of AGNs. \cite{delv2021} experienced a similar phenomenon: when they restricted their galaxy sample to only include sources with strong detections in the radio regime, they found that $\sim$16\%  of the population with high S/N radio observations were identified as radio AGNs.}

\begin{figure}
\includegraphics[width=\columnwidth]{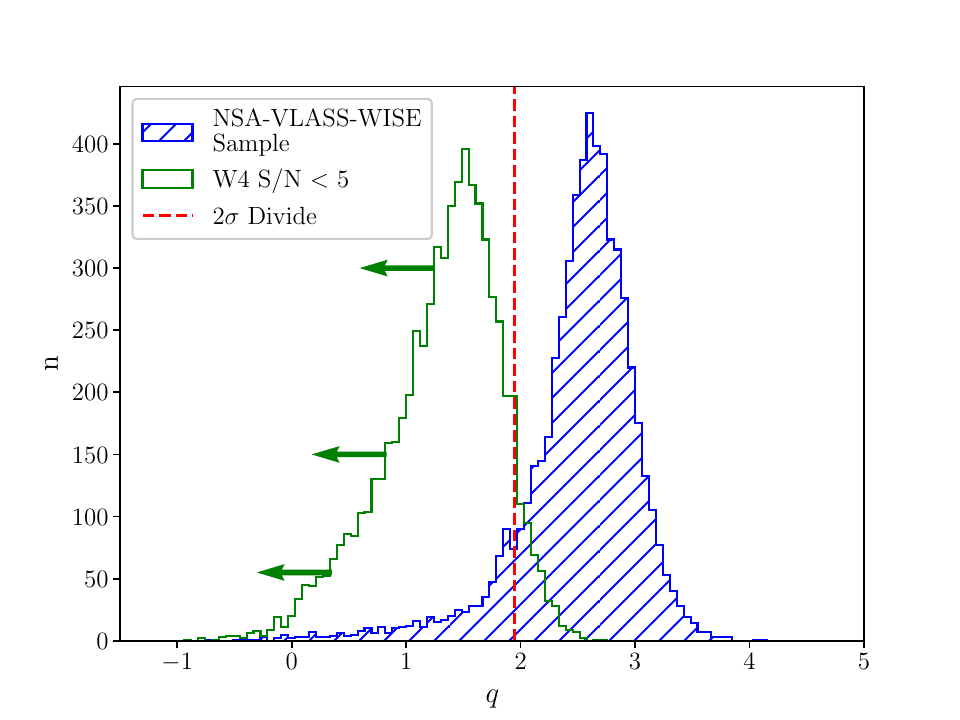}
 \caption{{Distribution of $q$ for the galaxies in the NSA-VLASS-WISE sample (blue) and for galaxies with WISE W4 observations with S/N $<$ 5 (green). The $q$ values calculated for the galaxies with W4 S/N $<$ 5 are upper limits, as indicated by the arrows. The 2$\sigma$ radio-excess cutoff derived from Figure \ref{fig:qdivide} is shown as a red dashed line. The vast majority of the galaxies with W4 S/N $<$ 5 have $q$ values inconsistent with SFGs and are dominated by a population of radio-excess AGNs.}}
\label{fig:q_sn_below5}
\end{figure}

{We also consider how the distribution of $q$ values may change with the inclusion of galaxies with W4 S/N $<$ 5. When we examine the distribution of the galaxies with W4 S/N $<$ 5, we find that it has much lower values of $q$ compared to the distribution of galaxies with high W4 S/N data (see Figure \ref{fig:q_sn_below5}). In fact, 93\% of the galaxies with W4 S/N $<$ 5 have $q$ $<$ 1.94 (our criterion established above) and are likely radio-excess AGNs. Moreover, this statistic does not consider the fact that the $q$ values calculated with W4 S/N $<$ 5 are upper limits, meaning that the true percentage of galaxies with $q$ $<$ 1.94 could be even higher. As such, the low S/N observations largely do not affect the distribution of the SFGs. Since the radio-excess AGN cutoff was established using a population dominated by SFGs, the effect that the W4 S/N $<$ 5 galaxies have on the radio-excess AGN cutoff is minimal.}

\section{Properties of the NSA-VLASS-WISE Sample}
\label{sec:Properties_NSA-VLASS-WISE}
\subsection{Radio Source Properties}
\label{subsec:radio_source_properties}
We consider the characteristics of all the radio sources in {the NSA-VLASS-WISE sample that are not mid-IR selected AGNs} and compare the properties of the galaxies selected as radio-excess AGN hosts to the remaining galaxies. For simplicity of nomenclature, we refer to all galaxies not identified as radio-excess AGN hosts as being SF-consistent, even though they could still contain AGNs, albeit AGNs that do not have radio emission greater than the emission predicted to come from SF. 

\begin{figure*}
 \includegraphics[width=\textwidth]{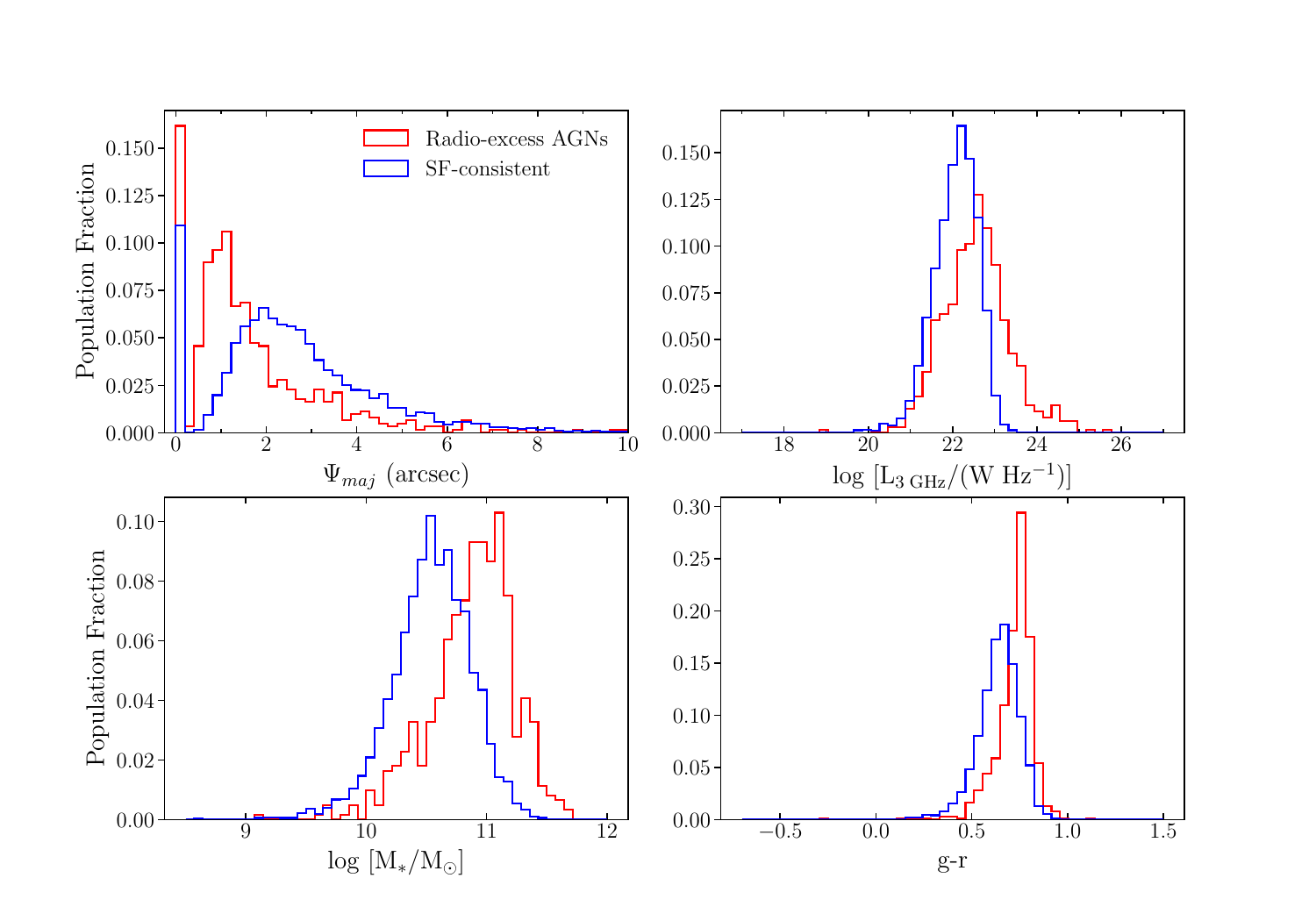}
 \caption{Histograms comparing the radio-excess AGN host population (red) to the SF-consistent population (blue) of the galaxies in the NSA-VLASS-WISE sample {that are not mid-IR AGNs}. Top left: Deconvolved major axis sizes, $\Psi_{maj}$. Values closer to zero are indicative of more compact sources. Top right: Integrated 3 GHz radio luminosities of the radio sources. Bottom left: Stellar masses of the host galaxies. Bottom right: $g-r$ colors of the host galaxies. In general, the radio-excess AGNs have radio sources that are more compact and host galaxies that are more massive and redder than the SF-consistent galaxies.} 
\label{fig:upper_lower_wing}
\end{figure*}

We examine the radio sources of the two populations to see if they are point-like by analyzing the distribution of the full width at half maximum (FWHM) of the deconvolved major axis ($\Psi_{maj}$) of the sources. VLASS can probe down to a size of $\sim$0\farcs3 and below this angular size, $\Psi_{maj}$ values are unresolved and are recorded in the VLASS catalog as having a value of 0\arcsec\ \citep{gordon2021}.
We create normalized histograms of $\Psi_{maj}$ values for the radio-excess AGNs and the SF-consistent population (see Figure \ref{fig:upper_lower_wing}).
We find that the fraction of sources with $\Psi_{maj}=0\arcsec$ is larger for the radio-excess AGN population (16\%) than the SF-consistent population (11\%).
Additionally, the AGN population contains a larger percentage of compact sources, with a median value of $\Psi_{maj}=$1\farcs2 while the SF-consistent population has a median value of $\Psi_{maj}=$2\farcs4 (both medians are calculated by ignoring the 0\arcsec\ values). 
A two-sample Kolmogorov-Smirnov (K-S) test returns a p-value of {$p < 0.001$}, indicating that the distributions of $\Psi_{maj}$ originate from different parent samples.  

For decades, radio galaxies have been divided into two major types: Fanaroff–Riley Class I (FR-I) and Fanaroff–Riley Class II (FR-II; \citealt{fanaroff1974}). {FR-I sources are core dominated and have radio emission that peaks near the center of their host galaxies and fades with distance. FR-II sources are dominated by their jets and lobes and brighten near the edge of their hosts}.
Given the relatively small angular size of the radio sources in the radio-excess AGN population, we hypothesize that our sample of radio-excess AGNs primarily consists of core-dominated sources. This result is expected, since in the local universe, the majority of radio galaxies are FR-I sources (\citealt{ledlow1995}; \citealt{govoni2000}; \citealt{best2005b}).
Furthermore, in cross-matching the VLASS and NSA catalogs, we required the radio sources to be located close to the center of their host galaxies, constraining our sample to preferentially detect FR-I sources.



We also compare the 3 GHz radio luminosities of the two populations (see Figure \ref{fig:upper_lower_wing}). The distribution of radio luminosities for the SF-consistent population has a median of log($\mathrm{L_{3\;GHz}}$/$\mathrm{{W\; Hz^{-1}}}$) = 22.1 and a FWHM of 1.3 
and the radio-excess AGN population has a median radio luminosity of log($\mathrm{L_{3\;GHz}}$/$\mathrm{{W\;Hz^{-1}}}$) = 22.6 with a FWHM of 1.9. 
A two-sample K-S test of the distributions reveals that the probability that the two distributions come from the same original distribution is extremely low, with a probability of {$p < 0.001$}.
We note that the two populations have significant overlap in radio luminosities despite being statistically different. 
This overlap reiterates the fact that radio luminosity alone is often not enough to determine if a source is a radio-excess AGN. {Using the definition of radio-quiet AGNs from \cite{miller1990} as AGNs with radio luminosities $L_{\nu} < 10^{24}$ W Hz$^{-1}$, we see that both the SF-consistent population and the radio-excess AGN population can host radio-quiet AGNs.}
{Despite the overlap between the two samples, the radio-excess AGN radio luminosity distribution has a prominent tail toward high luminosity which results in the small p-value of the two-sample K-S test.} The SF-consistent population has a maximum radio luminosity of log($\mathrm{L_{3\;GHz}}$/$\mathrm{{W\;Hz^{-1}}}$) = {23.9}, while the radio-excess AGN population has a maximum of log($\mathrm{L_{3\;GHz}}$/$\mathrm{{W\;Hz^{-1}}}$) = 26.0. As such, it could be hypothesized that a 3 GHz luminosity above $\sim10^{24} \;\mathrm{W\;Hz^{-1}}$ is indicative of a radio-loud AGN. 


\subsection{Host Galaxies}
\label{subsec:host_gals}
We compare the populations of galaxies by investigating their stellar masses and $g-r$ colors, taken from the NSA. Normalized histograms of the estimated stellar masses ($M_{*}$) of the two populations are shown in Figure \ref{fig:upper_lower_wing}. 
The mass distribution of SF-consistent population has a median stellar mass of log($M_{*}/M_\odot$) = 10.6 with a FWHM of 0.82 and the radio-excess AGN distribution has a median of log($M_{*}/M_\odot$) = 10.9 and a FHWM of 0.92. 
{The two sample K-S test returns a p-value of $p < 0.001$, indicating that the two distributions originate from totally different parent samples.}

The $g-r$ colors of the two populations, calculated from the absolute AB magnitudes in the NSA, are also shown in Figure \ref{fig:upper_lower_wing}. The SF-consistent population has a median color of $g-r=0.65$ with a FWHM of 0.40 and the radio-excess AGN population has a median of 0.75, with a FWHM of 0.25. The two-sample K-S test returns a p-value of $p < 0.001$, showing that the two distributions are distinct and are unlikely to have come from the same parent distribution. Overall, the population of radio-excess AGN hosts appears to be redder and more massive than the SF-consistent galaxies. 

\begin{figure}
\includegraphics[width=\columnwidth]{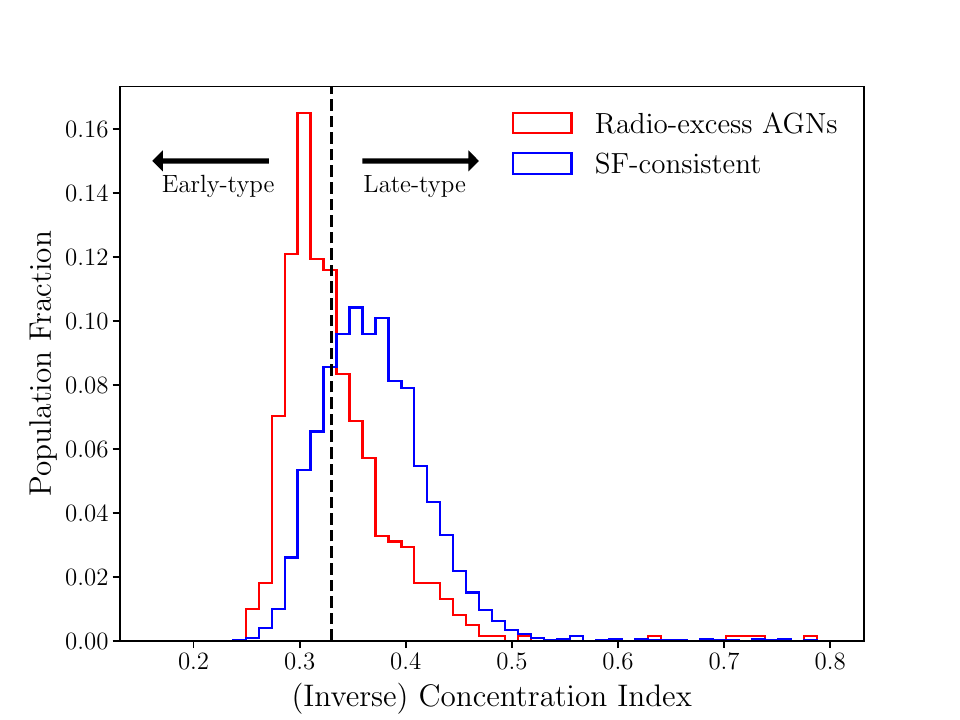}
 \caption{Histogram of the (inverse) concentration indices, defined as $r_{50}/r_{90}$, for the radio-excess AGN and SF-consistent populations in the NSA-VLASS-WISE sample.  The \cite{shimasaku2001} divide between late and early-type galaxies is shown as a dashed line. Notably, the radio-excess AGN population peaks in the early-type regime while the SF-consistent population peaks in the late-type regime.} 
\label{fig:inverse_conc_hist}
\end{figure}

We also determine that the SF-consistent galaxies tend to be late-type galaxies. We quantify this in Figure \ref{fig:inverse_conc_hist} by comparing the (inverse) concentration indices of the two populations, {calculated using the half-light and 90\%-light Petrosian radii from the NSA.} 
The (inverse) concentration index comes from \cite{shimasaku2001} and is defined  as $C=r_{50}/r_{90}$, where $r_{50}$ and $r_{90}$ are the half-light and 90\%-light radii, respectively. The divide between early and late type galaxies, as established by \cite{shimasaku2001}, is also shown, with the peak of the radio-excess AGN population being found on the early-type side of the divide and the peak of the SF-consistent population being found on the late-type side. {In addition to being visually distinct, the distributions are statistically different, with the two-sample K-S test returning a p-value under 0.001.}
{The (inverse) concentration indices also show that the radio-excess AGN hosts are more centrally concentrated than the SF-consistent galaxies. The compactness of the AGN hosts could be driven by the presence of the AGN as a central point source, although given that the radio properties of an AGN rarely correlate well with their optical properties (e.g, \citealt{best2005b}), we find this to be unlikely. It is more likely that AGNs reside in more centrally concentrated galaxies than the presence of a radio AGN influences the calculation of the (inverse) concentration indices.}

\begin{figure}
\includegraphics[width=\columnwidth]{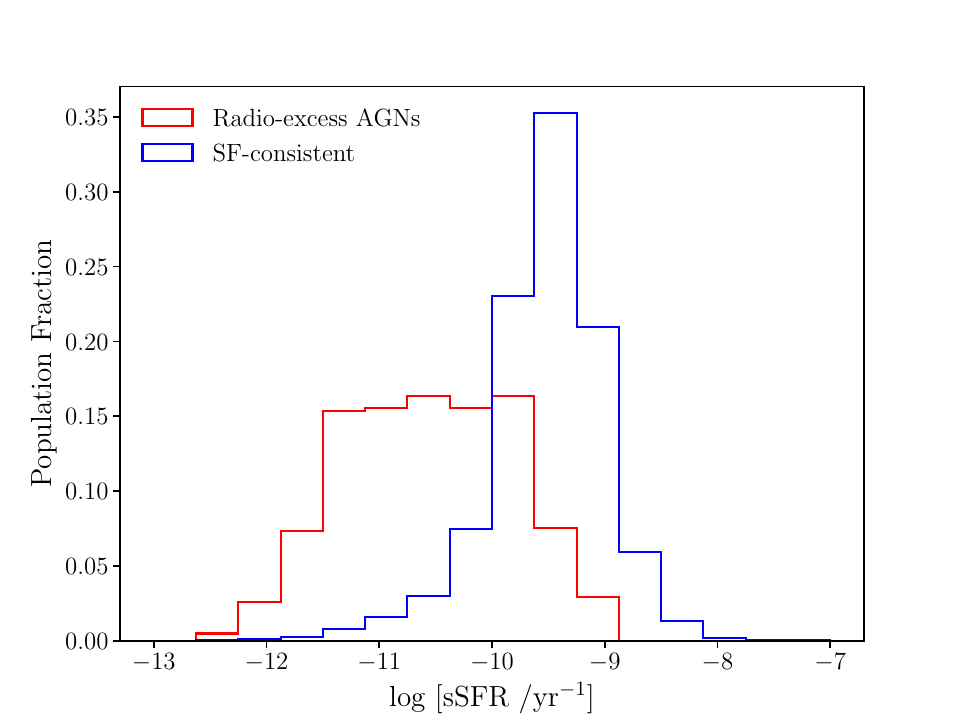}
 \caption{sSFRs for the radio-excess AGN and SF-consistent populations in the NSA-VLASS-WISE sample. The SF-consistent population predictably have higher sSFRs, while the radio-excess AGNs have a wider spread and a larger fraction of its population with lower values of sSFR. The low values of sSFR in the AGN population could be caused by SF quenching from the AGNs.} 
\label{fig:ssfr}
\end{figure}

We also consider the specific SFR (sSFR) of the galaxies to see if there is evidence of quenching of SF within the AGN hosts (see Figure \ref{fig:ssfr}).

{We calculate SFRs using IR data from WISE and the following SFR$_{IR}$ equation from \cite{murphy2011}}:
\begin{equation}
    \frac{\mathrm{SFR}_{IR}}{M_\odot \: yr^{-1}} = 3.88 \times 10^{-44} \left( \frac{\mathrm{L_{TIR}}}{erg \: s^{-1}} \right)
    \label{eq:sfr}
\end{equation}
where $\mathrm{L_{TIR}}$ is the TIR luminosity.
This equation was derived  by integrating the output spectrum from Starburst99 stellar population models \citep{leitherer1999} over the wavelength range $912 \;\mathrm{\AA} < \lambda < 3646\; \mathrm{\AA}$. The models used a Kroupa (\citealt{kroupa2001}) initial mass function (IMF), with a slope of -1.3 for masses between 0.1 and 0.5 \(M_\odot\) and -2.3 for masses between 0.5 and 100 \(M_\odot\).
We use the $\mathrm{L_{TIR}}$ values found in \S\ref{subsec:IRRC} to calculate the SFRs. 
We divide the SFRs by the stellar masses from the NSA to find the sSFRs.

The SF-consistent population has a sharper peak at higher sSFRs than the AGN population, with a median value of log(sSFR/$\mathrm{yr^{-1}}$) = -9.5 and FWHM of 1.2 compared to the AGN population's median of log(sSFR/$\mathrm{yr^{-1}}$) = -10.6 and FWHM of 1.7. {The sSFR distributions are also statistically different and the two-sample K-S test returns a p-value below 0.001.}
 
\begin{figure}
\includegraphics[width=\columnwidth]{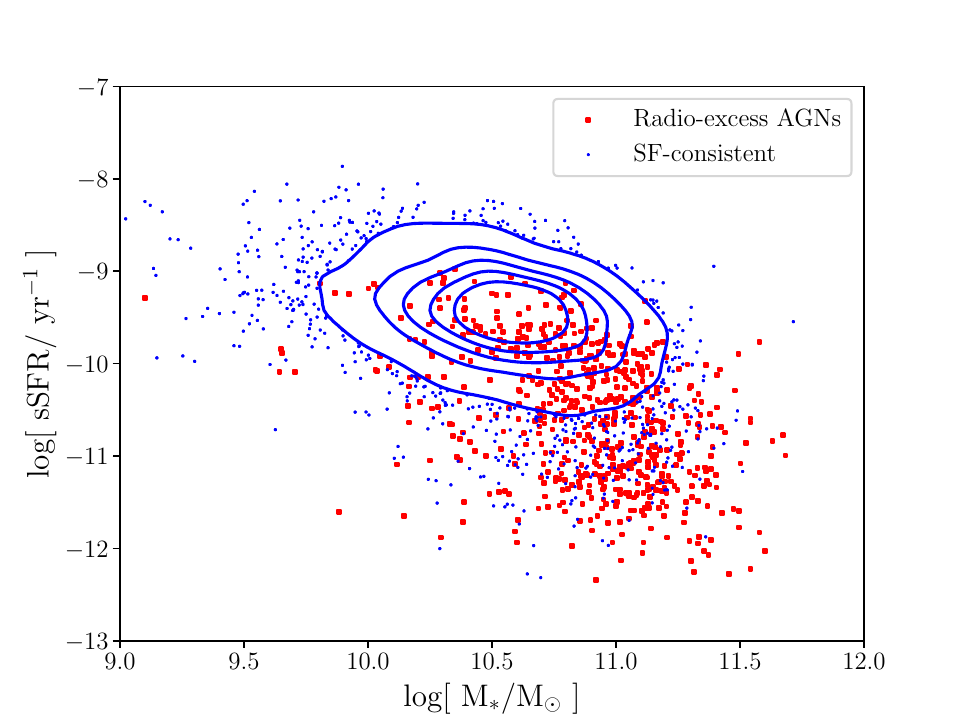}
 \caption{{sSFRs versus stellar masses for the radio-excess AGN hosts and SF-consistent populations in the NSA-VLASS-WISE sample. The blue contours and points show the distribution of the SF-consistent galaxies, and the red squares show the radio-excess AGN hosts. At stellar masses above $\sim10^{10.7}M_\odot$, the radio-excess AGN hosts have consistently lower sSFRs than the SF-consistent galaxies. Below this mass, however, the sSFRs of the radio-excess AGN hosts often have similar values to the sSFRs of the SF-consistent galaxies.}} 
\label{fig:ssfr_mass}
\end{figure}

Overall, the sSFR distributions show that the AGN population has a larger percentage of galaxies with lower amounts of SF per unit mass, which could be indicative of the AGN quenching SF in their host galaxies. {However, when we compare sSFRs to stellar masses (Figure \ref{fig:ssfr_mass}), we find that this is largely true only for larger mass galaxies. Radio-excess AGN host galaxies with stellar masses below $\sim10^{10.7}M_\odot$ are more likely to have sSFRs comparable to SF-consistent galaxies of similar mass. This suggests that the AGNs in smaller galaxies are less effective at quenching SF as their counterparts in more massive host galaxies.} 

\subsection{Comparison to the Literature}
The masses, colors, and morphological types of these populations agree well with the literature. It has been known for decades (\citealt{matthews1964})  that most radio AGNs are hosted by massive galaxies ($M_{*}\sim10^{11}M_\odot$), a result that has been verified in the local universe (up to $z$ $\sim$ 1) with direct estimates (e.g., \citealt{heeschen1970}; \citealt{ekers1973}; 
\citealt{jenkins1982}; \citealt{best2005b}; \citealt{mauch2007}; \citealt{smolcic2009}; \citealt{magliocchetti2016}; \citealt{sabater2019}; \citealt{capetti2022}). Our sample of radio-excess AGNs also largely resides in larger galaxies, with 37\% of the AGN hosts having masses $M_{*} > 10^{11}M_\odot$. 
Similarly, the fact that our sample of radio-excess AGNs has a redder color and resides in early-type galaxies matches canonical results. Research has established that the radio AGN host population is dominated by galaxies with older stellar populations (age $\sim$ 8-14 Gyr; e.g., \citealt{nolan2001}; \citealt{best2005b}) and galaxies that have a lack of continuing SF (e.g., \citealt{ledlow1995}; \citealt{govoni2000}; \citealt{siebenmorgen2004}; \citealt{dicken2012}). As such, radio AGNs have been shown to primarily reside in “red and dead” elliptical galaxies, which match the results we find in our sample.
For a thorough review of the general properties of radio AGNs and their host galaxies, see the \cite{maglio2022} review and references cited therein. 

\subsection{Mid-IR AGNs}
\label{subsec:midIR_AGNs_all}
In \S\ref{subsec:remove_midIR}, we removed mid-IR-selected AGNs to ensure that the distribution of $q$ was not significantly contaminated by IR emission from AGNs. We return to this sample of mid-IR AGNs to determine how many are also classified as radio-excess AGNs. Applying the $q < 1.94$ threshold, we find that of the {1,103} galaxies with mid-IR-selected AGNs, 98 ($\sim$9\%) are also selected as radio-excess AGNs. 
This matches well with canonical evidence that has shown that $\sim$$10-15$\% of AGNs are radio-loud (e.g., \citealt{kellermann1989}; \citealt{stern2000b}; see also the \citealt{urry1995} review). We note, however, that radio-loud AGNs are not exactly equivalent to radio-excess AGNs {since radio-loud AGNs are typically based on absolute radio luminosities or the ratio of radio to optical fluxes, whereas radio-excess AGNs are based on the ratio of radio to IR luminosities.}

\begin{figure}
\includegraphics[width=\columnwidth]{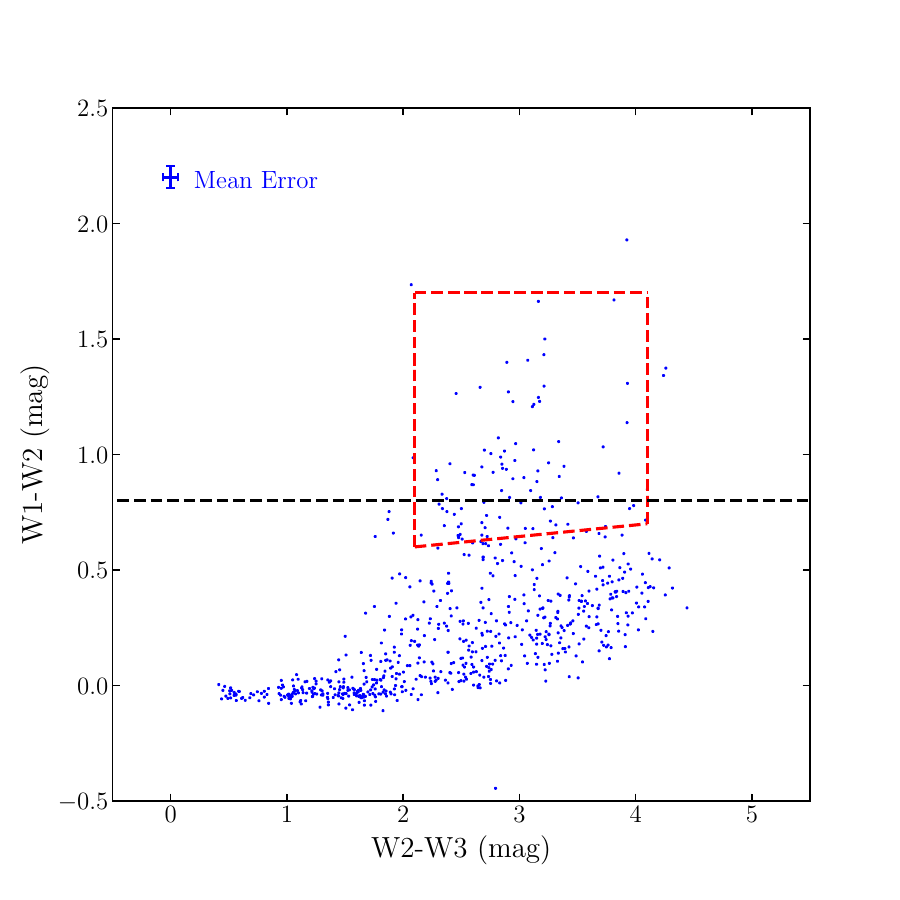}
 \caption{WISE color-color diagram for all galaxies in the NSA-VLASS-WISE sample identified as hosts for radio-excess AGNs. Approximately 14\% of the radio-excess AGNs are also classified as mid-IR AGNs, since they lie within the \cite{jarrett2011} selection box (shown in red) or above the \cite{stern2012} cutoff (shown in black). {The mean error bars for the data are shown in the upper left corner.}}
\label{fig:reas_wise_color}
\end{figure}

The mid-IR-selected AGNs also make up a significant percentage of all the radio-selected AGNs in the full NSA-VLASS-WISE sample. Of the {723} radio-excess AGNs identified, 98 (14\%) are also identified as mid-IR AGNs based on WISE color-color cuts (see Figure \ref{fig:reas_wise_color}). 


\section{Dwarf Galaxies with Radio-excess AGN\lowercase{s} in VLASS}
\label{sec:Dwarfs}
\subsection{Selection Criteria}
\label{subsec:select_dwarfs_agns}
Our primary goal of this work is to identify and study dwarf galaxies with radio-excess AGNs. We create a sample of dwarf galaxies from the NSA in a manner consistent with \cite{reines2020}, selecting galaxies with stellar masses $M_{*} \leq 3 \times 10^{9} M_\odot$, an upper mass limit approximately equal to the stellar mass of the Large Magellanic Cloud (LMC). Applying the mass restriction to the NSA results in a sample of 63,656 dwarf galaxies.

We cross-correlate our dwarf galaxy sample from the NSA catalog to the radio sources from VLASS, requiring a cross-matching of radius $\leq$ 2\farcs5 (the resolution of VLASS). We find 123 matches throughout the NSA volume that meet this criterion. We cross-match these galaxies to the ALLWISE catalog with a radius of 2\farcs5. {Unlike what was done with the full NSA-VLASS-WISE sample, we do not restrict by the WISE S/N for the dwarf galaxies for two reasons. First, we are not using the dwarf galaxies themselves to calibrate our method and second, dwarf galaxies with VLASS detections are quite rare and we want to consider as many objects as possible. Indeed, AGN selection methods derived from more massive galaxies are often applied to dwarf galaxies for similar reasons \citep[e.g.,][]{reines2013,hainline2016}. {We reiterate that the inclusion of galaxies with low WISE S/N does not dramatically affect the distribution of $q$ for SFGs or the radio-excess AGN cutoff, as explained in \S\ref{subsec:IRRC}. Therefore, the previously established radio-excess AGN cutoff is still valid for this sample of galaxies. }}


We visually inspect each of the 123
objects flagged as dwarf galaxies to eliminate interlopers. We first eliminate 7
sources that are nearby HII regions. We further remove 14 objects that are distant quasars. These sources have the distinct blue, point-like appearance of QSOs and have redshifts in SDSS that are larger than the cutoff of $z \le 0.15$ employed by the NSA. We also remove 1 galaxy with a reported mass of 0 in the NSA, leaving 101 objects.

We inspect the spectra of the remaining galaxies to verify the redshift reported by the NSA, since galaxies with unreliable redshifts will also have unreliable stellar mass estimates. {We discard any galaxies whose spectra cannot be visualized in either SDSS or the NASA/IPAC Extragalactic Database (NED). Using the spectra from SDSS and NED, we calculate the redshifts for each of the galaxies and compare these to the redshifts reported in the NSA. If there is a discrepancy indicating the NSA redshift is erroneous, we remove the corresponding galaxy from our sample.}
Of the 101 galaxies, 50 have spectra in SDSS and/or NED with reliable fits that match the redshifts given by the NSA. We consider these galaxies to be bona fide dwarf galaxies.


\begin{figure*}[t]
\includegraphics[width=\textwidth]{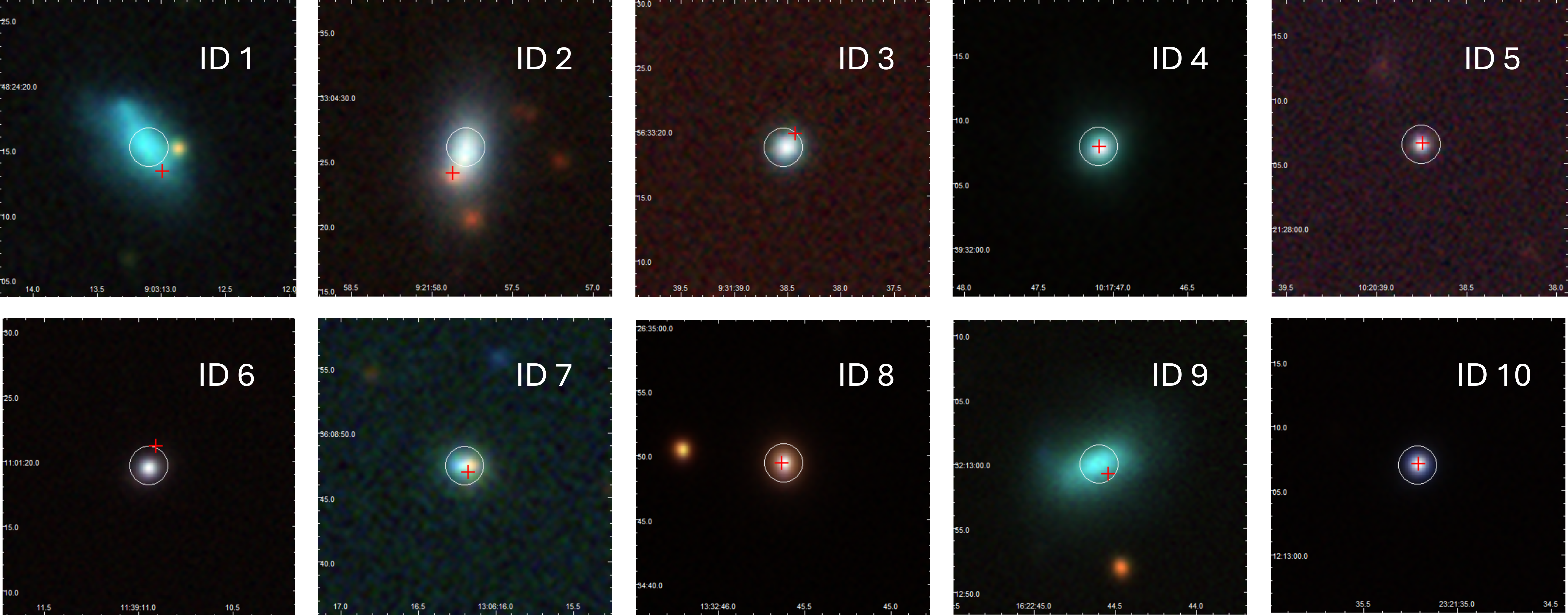}
 \caption{
 $grz$-band DECaLS images of the dwarf galaxies that are strong candidates for hosting radio-excess AGNs. The positions of the SDSS spectroscopic fibers with radii of 1.5\arcsec are shown as a white circles and the locations of the radio sources are shown as red crosses.} 
\label{fig:decals_pics}
\end{figure*}

\begin{figure*}[]
\includegraphics[width=\textwidth]{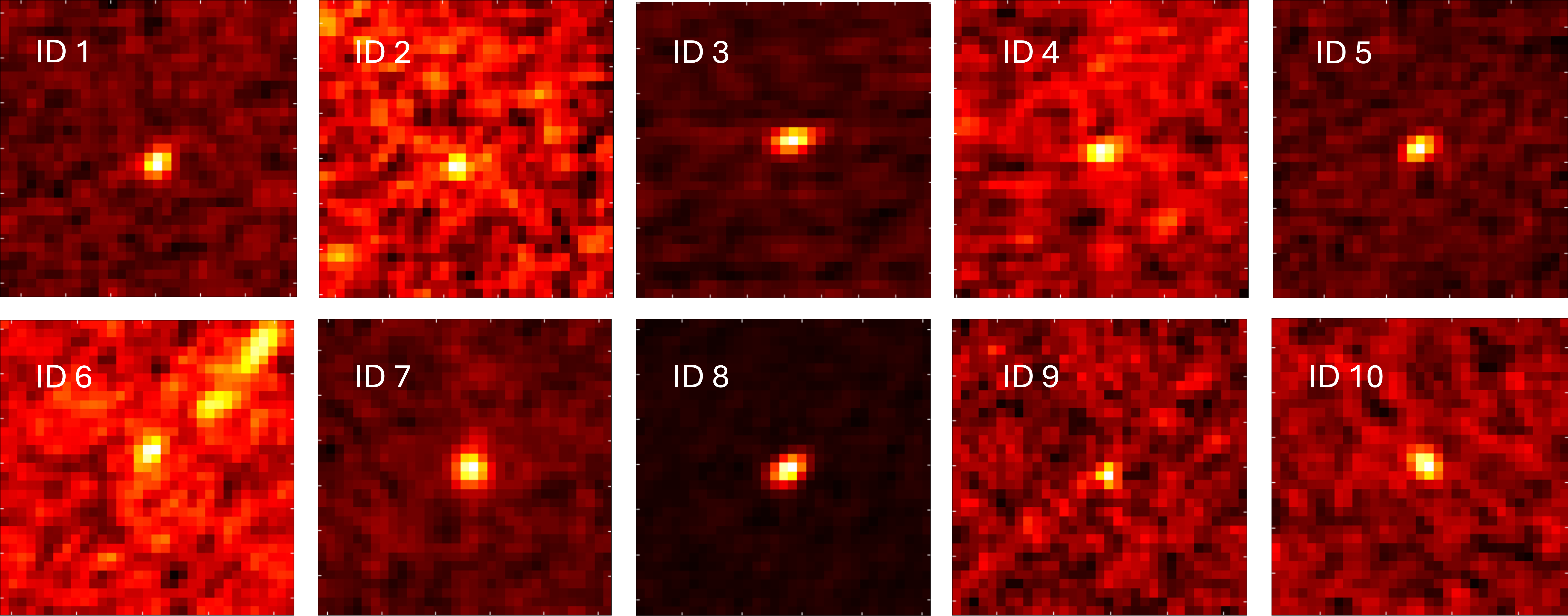}
 \caption{
 {VLASS images of the dwarf galaxies that are strong candidates for hosting radio-excess AGNs. The images are the same angular size as the DECaLS images in Figure \ref{fig:decals_pics}.}} 
\label{fig:vlass_pics}
\end{figure*}

We apply the $q < 1.94$ cutoff derived in \S\ref{subsec:IRRC} to the bona fide dwarf galaxies and find that 12 are inconsistent with SF, making them strong candidates for hosting radio-excess AGNs.
{Ideally we would have enough statistical power to calibrate the selection in the mass range of interest, but this is not the case due to the small number of dwarf galaxies in VLASS. However, we note that applying the same $q$ cutoff established from higher mass galaxies to dwarf galaxies is a very conservative method for finding AGN candidates. \cite{delv2021} used a sample of over 400,000 galaxies to show that the average value of $q$ increases with decreasing stellar mass. When we divide our sample into the same stellar mass bins used by \cite{delv2021}, we also find that the average value of $q$ increases as stellar mass decreases. While we consider extrapolating the selection criterion down to dwarf galaxy masses, we find considerable uncertainty in the average value of $q$ for galaxies with masses $M_{*} \leq 10^{9.5} M_\odot$ due to the small sample size in this mass range. Thus, we take a conservative approach and use the $q$ cutoff derived from higher mass galaxies even though the actual cutoff value for dwarf galaxies is likely to be higher. In doing so, we may lose potential AGNs, but we create a cleaner sample of AGNs in dwarf galaxies that have excessively low $q$ values.}



We remove 2 of the 12 objects inconsistent with SF from our sample, since they are known from other studies to {\it not} be AGNs associated with the dwarf host galaxies, leaving us with a final sample of 10 strong AGN candidates in dwarf galaxies.
One of the two objects that we remove is the radio source in J1313+4717, which is known to host a radio-loud SNe. \cite{corsi2014}, \cite{palliyaguru2019}, and \cite{palliyaguru2021} identified the radio source in J1313+4717 as a radio-loud type Ic SN, known as PTF11qcj. PTF11qcj is a rare type of SN, since most Ic SNe are radio-quiet, with radio luminosities $L_{\nu} < 10^{19} \mathrm{\;W\; Hz^{-1}}$ (\citealt{berger2003}; \citealt{soderberg2006}; \citealt{corsi2016}).
Due to its large radio luminosity {of $\mathrm{L}_{3\;\mathrm{GHz}} \approx 6 \times 10^{21} \mathrm{\;W\; Hz^{-1}}$}, it failed to be categorized as SF-consistent by the IRRC. 
Yet, given the extensive, multi-wavelength analysis performed on this object and its light curves, a SN is the most probable explanation for this radio source, not an AGN.
The other object that we remove is the radio source in J1136$+$1252. This source was identified as a radio AGN in a dwarf galaxy by \cite{reines2020} (ID 64 in their paper), but follow-up spectroscopy of the bright optical point source coincident with the off-nuclear radio source has revealed it to be a background quasar at $z=0.761$ (Sturm et al.,\ in preparation). {This source was not discarded previously because the redshift of the dwarf galaxy is correct. Only with our follow-up spectroscopy did we determine the radio source had a much higher redshift.}

{We provide Dark Energy Camera Legacy Survey (DECaLS) $grz$-band images of the 10 dwarf galaxies in our final sample in Figure \ref{fig:decals_pics}} {and provide VLASS images in Figure \ref{fig:vlass_pics}.} For ease of reference, we assign each galaxy an ID number.

\begin{figure}
 \includegraphics[width=\columnwidth]{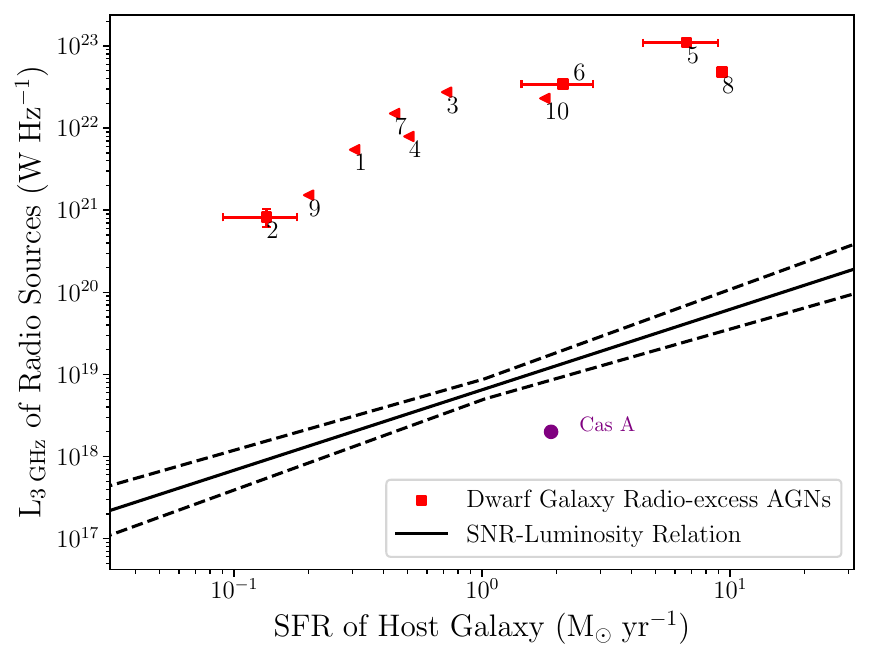}
 \caption{Radio source luminosity versus SFR of the 10 sources identified as radio-excess AGNs in dwarf galaxies. Each source is labeled with its respective ID number (see Figure \ref{fig:decals_pics} and Table \ref{tab:dwarf_agns}). The expected relation between SFR and radio luminosity of the brightest individual SNR/SNe from \cite{chomiuk2009} is shown as a solid black line, and the dashed lines show the expected scatter due to random statistical sampling. The SNR Cas A is shown for reference. {The error bars for the SFR calculations are shown, and galaxies with upper limit calculations of SFR are shown as triangles. The 3 GHz luminosity error bars are smaller than the marker size and are not visible on the plot. } 
 } 
\label{fig:dwarf_snr_lum}
\end{figure}

As consistency checks, we also carry out the analysis in \cite{reines2020} and determine whether the remaining 10 radio sources can plausibly be explained by SNRs/SNe.  
We first use the relation between the radio luminosities of the brightest SNRs in a galaxy and the SFR of a galaxy given by \cite{chomiuk2009}:
\begin{equation}
  \frac{  \mathrm{L}_{1.4\;\mathrm{GHz}}^{\mathrm{max}} }{10^{24} \; \mathrm{erg \; s^{-1} \; Hz^{-1}} } = \left( 95_{-23}^{+31} \right) \times \frac{\mathrm{SFR^{0.98\pm0.12}}}{M_\odot \; \mathrm{yr^{-1}}}
    \label{eq_lum_sfr}
\end{equation}
We convert this relation to the radio luminosities at 3 GHz using a spectral index of $\alpha=-0.5$ \citep{chomiuk2009} and compare the expected luminosity from SNRs to the actual 3 GHz luminosities of the 10 radio-excess AGNs in Figure \ref{fig:dwarf_snr_lum}. We employ the SFRs calculated from Equation \ref{eq:sfr} for each galaxy. 
We find that the radio sources lie far above the maximum expected luminosities from individual SNRs from their galaxies, with all the sources being at least one magnitude larger than expected. As such, individual SNRs are not a likely explanation for the radio sources.
For comparison, we also include Cas A, one of the youngest and most luminous SNRs in the Milky Way. 
{Using a 1 GHz flux density of $\mathrm{S}_{1\;\mathrm{GHz}}$ = 2723 Jy and a spectral index of $\alpha$ = -0.770 from \cite{baars1977} (epoch 1980.0), and distance of 3.4 kpc from \cite{reed1995}, we calculate a 5 GHz spectral luminosity of $\mathrm{L}_{9\;\mathrm{GHz}} \sim 2 \times 10^{18} \; \mathrm{W\;Hz^{-1}}$ for Cas A. }

We also consider entire populations of SNRs as a possible source for compact radio emission by using the luminosity function derived in \cite{chomiuk2009}:
\begin{equation}
  n(L) = \frac{dN}{dL} = 92 \times \mathrm{SFR} \times L^{-2.07}
    \label{eq_lum_sfr_snrs}
\end{equation}
where $n(L)$ is the number of SNRs with 1.4 GHz luminosity $L$. We again convert from 1.4 GHz to 3 GHz using a spectral index of $\alpha=-0.5$ \citep{chomiuk2009} and calculate SFR using Equation \ref{eq:sfr}.

\begin{figure}
 \includegraphics[width=\columnwidth]{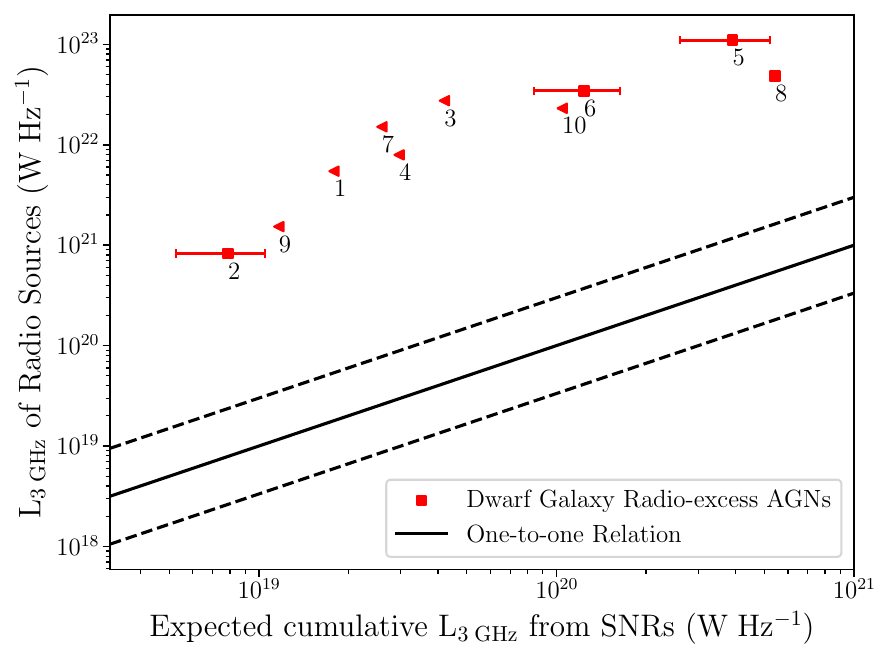}
 \caption{Observed 3 GHz luminosities versus the predicted cumulative radio luminosities from populations of SNRs/SNe for the 10 sources identified as radio-excess AGNs in dwarf galaxies. Each source is labeled with its respective ID number. The solid black line shows the one-to-one relation, with dashed black lines offset by a factor of three. {Galaxies whose predicted cumulative luminosities are upper limits (based on Equation \ref{eq_lum_integral}) are shown as triangles, the other galaxies have error bars shown. The 3 GHz luminosity error bars are smaller than the marker size and are not visible on the plot.} 
} 

\label{fig:dwarf_total_snr_lum}
\end{figure}

The total expected luminosity from all SNRs in a galaxy is calculated by taking the following integral from $10^{16} \; \mathrm{W \;Hz^{-1}}$ to $10^{21} \; \mathrm{W \;Hz^{-1}}$
to cover the range of SNRs/SNe presented in \cite{chomiuk2009}:

\begin{equation}
L_{\mathrm{total}} = \int n(L) \; L \; dL
    \label{eq_lum_integral}
\end{equation}
We calculate this integral and compare the total radio luminosity from SNRs to the observed 3 GHz luminosity in Figure \ref{fig:dwarf_total_snr_lum}. As with the individual SNR plot, we find the actual radio luminosities to dwarf the expected radio luminosities from the SNR population, leaving it unlikely that these phenomena are the source of the radio emission.

\subsection{Potential Background Contamination}
\label{subsubsec:dwarfs_background}
\begin{figure}
 \includegraphics[width=\columnwidth]{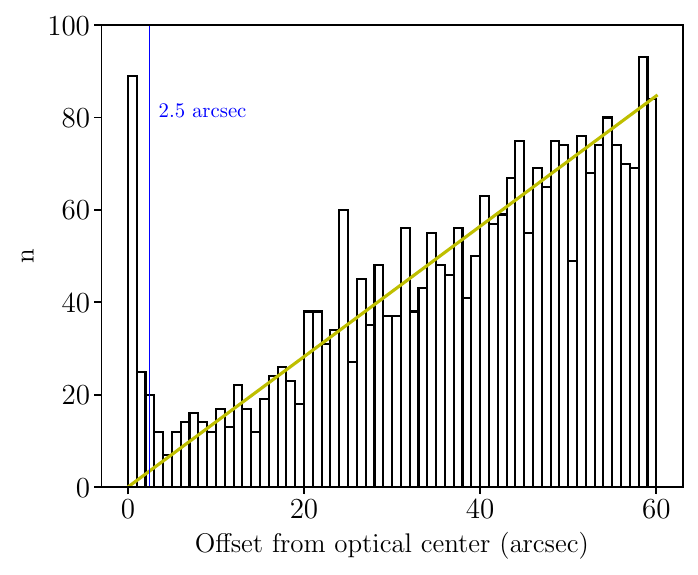}
 \caption{Observed offset distribution from cross-matching the NSA parent sample of dwarf galaxies with the VLASS radio sources, out to a matching radius of 60\arcsec. The best linear fit of the histogram is shown in yellow, and the actual cross-matching radius used to find radio sources in galaxies is shown in blue.}
\label{fig:histo60}
\end{figure}

Given the small cross-matching radius of 2\farcs5 utilized to identify VLASS sources within dwarf galaxies, we expect the percentage of background sources to be small. We verify this hypothesis using a method like that performed in \cite{reines2020}. We estimate the number of coincidental matches found for the whole sample of dwarf galaxies from the NSA by cross-matching the VLASS sources to the NSA galaxies out to a radius of 60\arcsec\ to find the offset distribution of coincidental matches. The offset probability histogram for background sources should be a Rayleigh distribution, equal to zero at an offset of zero and rising linearly for small offsets. We fit the offset probability histogram at radii greater than 2\farcs5 (to ignore the bias imposed by the large number of non-coincidental matches found below 2\farcs5) and find the offset probability function with respect to cross-matching distance (see Figure \ref{fig:histo60}). We integrate under the fit out to 2\farcs5 and find an expected number of $4.4 \pm \sqrt{4.4} \approx 4 \pm 2$ coincidental matches. Consequently, we expect less than 5\% of the 123 radio sources to be coincidental matches between dwarf galaxies and background radio sources, meaning that statistically, less than 1 of the 10 dwarf galaxies with AGNs is estimated to be a background source. 

\subsection{General Properties}
\label{subsubsec:dwarfs_general_props}
As was done with the NSA-VLASS-WISE sample, we consider several properties of the 10 radio-excess AGNs (see Table \ref{tab:dwarf_agns}) to see if there are any defining characteristics of radio-excess AGNs in dwarf galaxies.  We also contrast the properties of the dwarf galaxy AGNs to the AGNs in the full NSA-VLASS-WISE sample (see \S\ref{sec:Properties_NSA-VLASS-WISE}).

\begin{figure}
\includegraphics[width=\columnwidth]{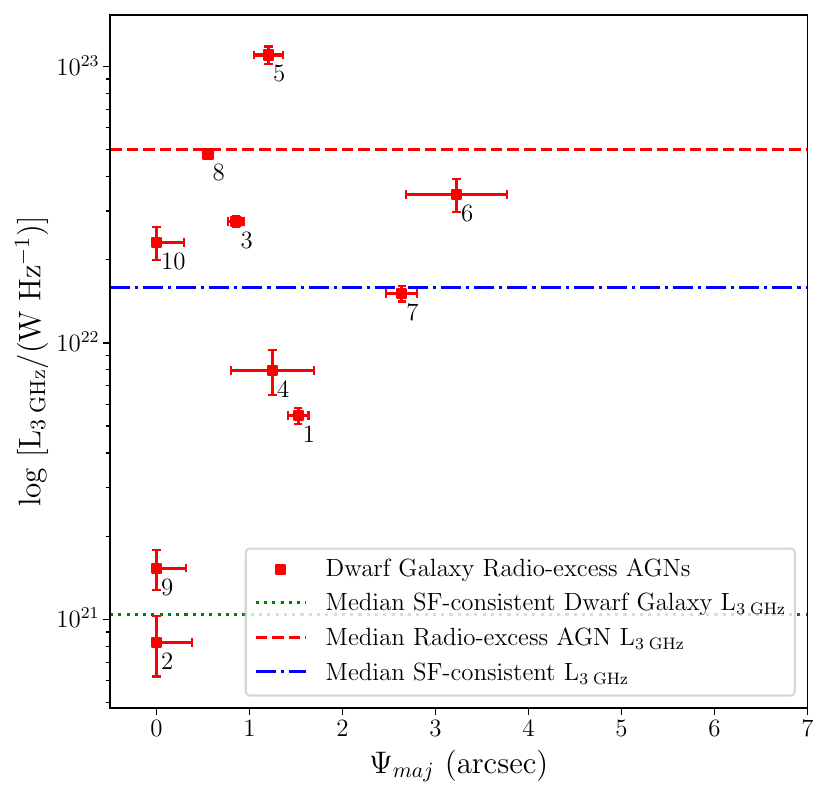}
 \caption{3 GHz luminosities versus deconvolved major axis sizes of the sources consistent with radio-excess AGNs in dwarf galaxies. The ID numbers for each galaxy are shown, {as well as the median value of radio luminosity for the SF-consistent dwarf galaxies (green dotted line).} The median values of the radio-excess AGNs and SF-consistent galaxies in the NSA-VLASS-WISE sample are also shown (red and blue lines, respectively). On average, the radio sources in the dwarf galaxies are dimmer than the radio sources of the AGNs in more massive galaxies.} 
\label{fig:dwarf_lum_dcmaj}
\end{figure}

We consider the features of the radio sources in Figure \ref{fig:dwarf_lum_dcmaj} by plotting the 3 GHz luminosity against the size of the deconvolved major axis of each source. {Following the standard employed by \cite{gordon2021}, we find that two of the dwarf galaxy radio-excess AGNs have major axis sizes $\Psi_{maj} > 2\farcs5$ and are considered to be extended sources.} {These large major axis sizes could be explained by the AGNs having jet and core morphologies. ID 6 in particular appears to have jet-like features in its VLASS image (Figure \ref{fig:vlass_pics}). The angular size of the radio sources could also be explained by the galaxies having their synchrotron emission dominated by SF. While we utilized the IRRC to eliminate SF as the exclusive cause of the radio emission, there could be a combination of SF and AGN emission, resulting in an extended radio source. These objects would be similar to the galaxies found by \cite{gim2019}, which 
had low $q$ values and signs of starburst activity, indicating the presence of both AGNs and SF.}
 The remaining sources have smaller angular sizes and are therefore more consistent with {compact sources}, which is what we expect for core-dominant AGNs and is what was seen for the full NSA-VLASS-WISE sample (see Figure \ref{fig:upper_lower_wing}).

In contrast, the radio luminosity of the dwarf galaxy sample are inconsistent with the full AGN sample, with the radio sources in the dwarf galaxies being much dimmer than the AGNs in the full sample. Only 1 source (ID 5) lies above the median radio luminosity of the full radio-excess AGN population, and 5 of the 10 sources have 3 GHz luminosities lower than the median radio luminosity of the full SF-consistent population. Assuming the radio emission comes from AGNs, this result is not surprising, since smaller galaxies are more likely to have smaller BHs that create less luminous emission. {We find that the median luminosity of the SF-consistent dwarf galaxies (green dotted line in Figure 14) is much lower than the median luminosities of the full population (blue dash-dot line).}  

\begin{figure}
\includegraphics[width=\columnwidth]{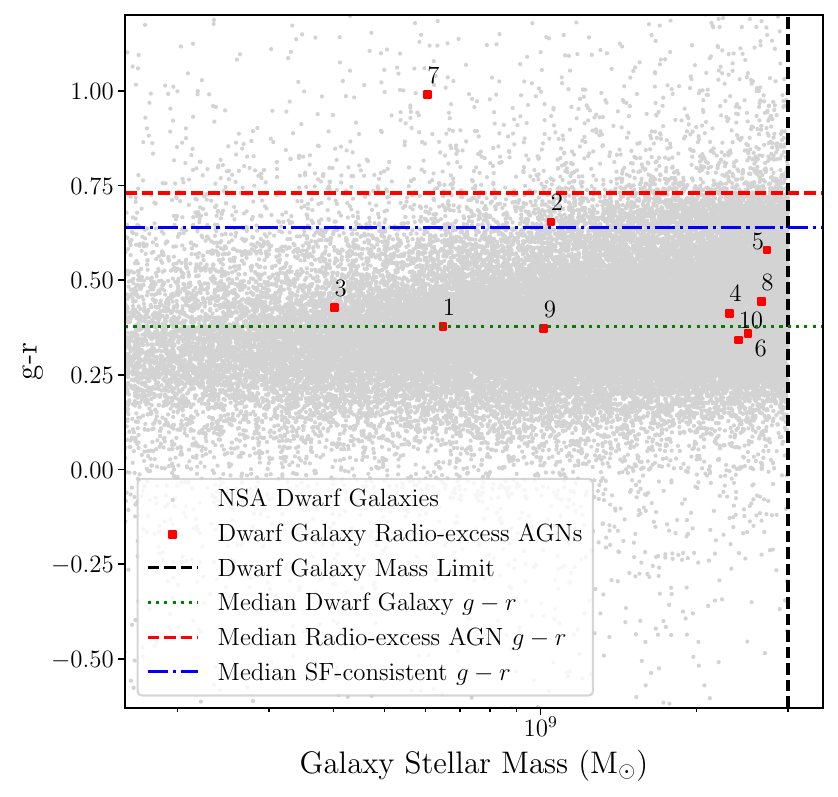}
 \caption{$g-r$ color vs. total stellar mass for dwarf galaxies ($M_{*} < 3\times10^9 M_{\odot}$) in the NSA. The AGN hosts, with their associated ID numbers are shown. They mostly follow the trend of the full NSA population. {The median color of all the dwarf galaxies in the NSA is shown as a green dotted line.} The median colors of the radio-excess AGNs and SF-consistent galaxies in the NSA-VLASS-WISE sample are also shown, indicating that the dwarf galaxies (including the AGN hosts) are much bluer than the more massive galaxies.} 
\label{fig:dwarf_mass_gr}
\end{figure}

We consider the properties of the AGN hosts by comparing the $g-r$ colors of the galaxies to their stellar masses in Figure \ref{fig:dwarf_mass_gr}. We also compare the galaxies to the full sample of dwarf galaxies in the NSA. We find that the radio-excess AGN hosts mostly follow the trend of the entire dwarf galaxy population, making it clear that color is not a clear indicator of AGN presence in a dwarf galaxy.
 
We also note that only 2 of the dwarf galaxies with radio-excess AGNs have $g-r$ colors that are redder than the median $g-r$ color of the full SF-consistent population and only 1 has a redder color than the median color of the full radio-excess AGN population. This is not surprising: dwarf galaxies are bluer in general than more massive galaxies. However, this does reiterate the fact that radio-excess AGNs in dwarf galaxies do not follow the trend of more massive galaxies of being found in redder galaxies. 




{The bluer colors of the dwarf galaxies hosting radio-excess AGNs could be explained by the fact that the AGNs in dwarf galaxies are not quenching SF as much as the AGNs in more massive galaxies. There is even the possibility that the AGNs are triggering SF in the dwarf galaxies, as was the case for the dwarf galaxy Henize 2-10 \citep{schutte2022}.}


\begin{figure}
\includegraphics[width=\columnwidth]{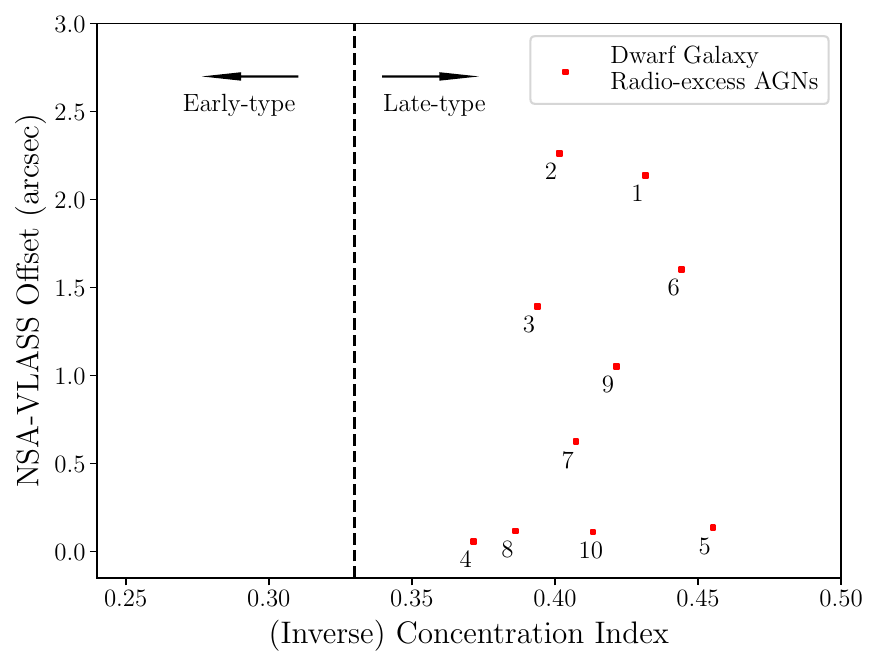}
 \caption{Offset of the radio sources in VLASS from the optical centers of their host galaxies versus the (inverse) concentration indices, defined as $C = r_{50}/r_{90}$ \citep{shimasaku2001}. Each point is labeled with the ID number of the respective galaxy. The dwarf galaxy AGN hosts have larger $C$ values than the full population of AGN hosts (Figure \ref{fig:inverse_conc_hist}), indicating that the dwarf galaxy hosts are less centrally concentrated than their more massive counterparts.
 } 
\label{fig:dwarf_offset_conc}
\end{figure}

\begin{deluxetable*}{cccccccccccccc}
\rotate
\tablenum{1}
\tablecaption{Radio-excess AGNs in Dwarf Galaxies}
\tablewidth{0pt}
\tablehead{
\colhead{ID} & \colhead{Name} & \colhead{NSA ID} & \colhead{R.A.} &\colhead{Dec.} & \colhead{$z$} & \colhead{log(M$_{*}$)} & \colhead{$g-r$} &  \colhead{Offset} & \colhead{S$_{3\;\mathrm{GHz}}$} & \colhead{log(L$_{3\;\mathrm{GHz}}$)} & \colhead{$\Psi_{maj}$} & \colhead{log(L$_{1.4\;\mathrm{GHz}}$)} & \colhead{log(L$_{\mathrm{TIR}}$)}}
\colnumbers
\startdata
  1 & J090313.09+482415.1 & 131284 & 135.8046 & 48.40421 & 0.0272 & 8.81 & 0.38  & 2.14  & 3.68(0.24)  & 21.74(0.03) & 1.5(0.1) & 21.99(0.03) & $<$35.90\\
  2 & J092157.79+330425.6 & 318231 & 140.4908 & 33.0739  & 0.0215 & 9.02 & 0.65  & 2.26 & 0.89(0.22)     & 20.92(0.11) & 0(0.4) & 21.15(0.11) & 35.54(0.14)  \\
  3 & J093138.51+563318.8 & 79697  & 142.9106 & 56.55521 & 0.0494 & 8.60 & 0.43   & 1.39 & 5.67(0.25)     & 22.44(0.02)  & 0.9(0.1)  & 22.67(0.02) & $<$36.27\\
  4 & J101747.08+393207.8 & 269031 & 154.4462 & 39.53552 & 0.0541 & 9.36 & 0.41   & 0.06 & 1.37(0.25)     & 21.90(0.08)  & 1.2(0.4) & 22.13(0.08) & $<$36.12\\
  5 & J102038.74+212806.5 & 500578 & 155.1615 & 21.46847 & 0.1370 & 9.44 & 0.58   & 0.14 & 3.02(0.22)     & 23.04(0.03)  & 1.2(0.2) & 23.29(0.03) & 37.24(0.14) \\
  6 & J113910.94+110119.5 & 236426 & 174.7956 & 11.02212 & 0.0837 & 9.40 & 0.36   & 1.60 & 2.50(0.34)     & 22.54(0.06) & 3.2(0.5) & 22.77(0.06) & 36.74(0.14)\\
  7 & J130616.20+360847.4 & 581178 & 196.5675 & 36.14651 & 0.0374 & 8.78 & 0.99   & 0.62 & 5.42(0.35)     & 22.18(0.03) & 2.6(0.2) & 22.40(0.03) & $<$36.06 \\
  8 & J133245.62+263449.3 & 484370 & 203.1901 & 26.58038 & 0.0469 & 9.43 & 0.44   & 0.12 & 10.94(0.26)    & 22.68(0.01) & 0.6(0.0) & 22.96(0.01) & 37.38(0.02)\\
  9 & J162244.60+321300.0 & 342650 & 245.6858 & 32.21666 & 0.0224 & 9.01 & 0.37   & 1.05 & 1.53(0.25)     & 21.18(0.07) & 0(0.3)   & 21.27(0.07) & $<$35.71\\
  10 & J232135.20+121306.9 & 606795 & 350.3967 & 12.2186  & 0.0720 & 9.38 & 0.34  & 0.11 & 2.25(0.30) & 22.36(0.06) & 0(0.3) & 22.60(0.06) & $<$36.66   
\enddata
\tablecomments{Column 1: galaxy ID. Column 2: galaxy name. Column 3: NSA ID (version v1\textunderscore0\textunderscore1). Column 4: R.A. of galaxy from the NSA, in units of degrees. Column 5: dec. of galaxy from the NSA, in units of degrees. Column 6: redshift given by NSA. Column 7: log galaxy mass given by the NSA, in units of $M_\odot$. Column 8: $g - r$ color given by the NSA. Column 9: offset between the galactic center given by NSA and the radio source in VLASS, in units of arcsec. Column 10: integrated radio flux density of radio source from VLASS, in units of mJy at 3 GHz. Column 11: log 3 GHz luminosity of radio source from VLASS, in W Hz$^{-1}$. Column 12: deconvolved major axis of radio source from VLASS, in arcsec. Column 13: log 1.4 GHz luminosity of radio source from FIRST, in units of W Hz$^{-1}$. Galaxies without FIRST detections have their radio luminosities estimated using a spectral index of $\alpha=-0.7$. Column 14: TIR luminosity, estimated using W4 magnitudes from WISE, in units of W. Upper limits are indicated. The final two columns are used to calculate $q$. Errors for the last 5 columns are given in parentheses. }
\end{deluxetable*}
\label{tab:dwarf_agns}
Further differences between the full radio-excess AGN population and dwarf galaxy radio-excess population are found by analyzing the (inverse) concentration indices $C$ (see Figure \ref{fig:dwarf_offset_conc}). Unlike the full population of radio-excess AGNs, which primarily have $C$ values consistent with early-type galaxies (Figure \ref{fig:inverse_conc_hist}), all of the dwarf galaxy radio-excess AGNs have $C$ values indicative of late-type galaxies. We note that the classification as early-type and late-type by the index $C$ does not perfectly match the classification of the dwarf galaxies by eye, since some of the galaxies appear irregular. While the classifications may not be accurate, the fact that the $C$ values are larger for the dwarf galaxies shows that the dwarf galaxies are less centrally concentrated than the AGN hosts of larger mass. 

We note that when \cite{reines2020} made a similar plot of offsets versus $C$ values for their sample of dwarf galaxy AGNs, there was a correlation between the two: higher offsets were more likely to have higher (inverse) concentration indices, seemingly indicating that more extended galaxies were more likely to host wandering AGNs. There is no clear correlation in our sample, however. If such a correlation only becomes clear at larger offsets, the maximum offset of 2\farcs5 employed in this paper may not be large enough to make the relation discernible. 

\begin{figure}
\includegraphics[width=\columnwidth]{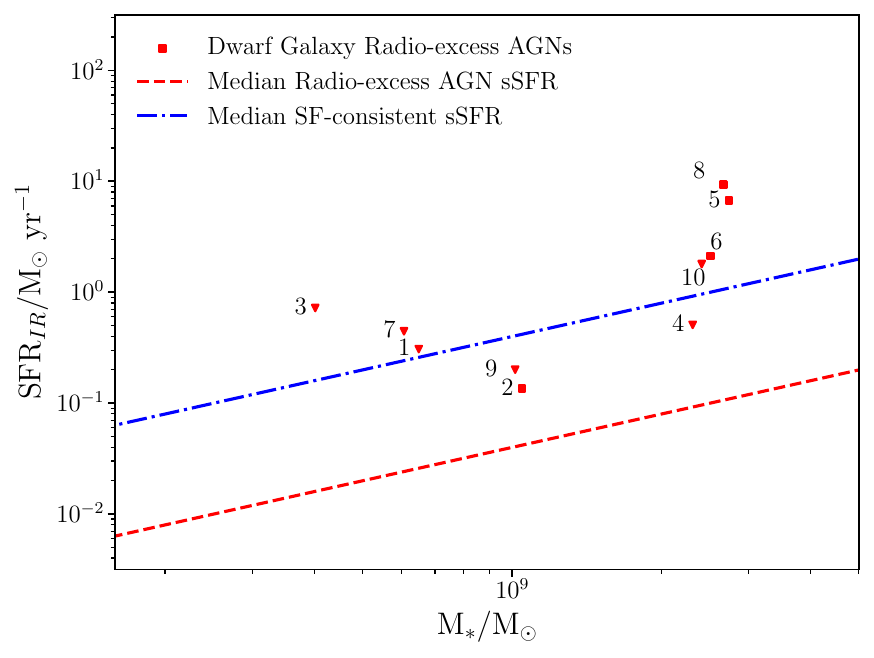}
\caption{SFR$_{IR}$ vs. stellar mass for the dwarf galaxies with radio-excess AGNs. SFRs calculated using upper limits in WISE are shown as triangles. The median sSFRs of the radio-excess AGN and SF-consistent populations for the full NSA-VLASS-WISE sample are also shown. In general, the dwarf galaxies have higher sSFRs than the full sample of galaxies. Each point is labeled with the ID number of the respective galaxy.} 
\label{fig:dwarf_mass_sfr}
\end{figure}

Additionally, we compare the sSFRs of the radio-excess AGNs in dwarf galaxies to the sSFRs of the full NSA-VLASS-WISE sample by plotting the SFRs against stellar masses (see Figure \ref{fig:dwarf_mass_sfr}). 
We find that all of the radio-excess AGNs in dwarf galaxies have sSFRs higher than the median sSFR of the full AGN sample. The majority of the galaxies have sSFRs that lie above the median values of the SF-consistent population, indicating that, in general, the dwarf galaxies (including those hosting radio-excess AGNs) have larger amounts of SF than the typical galaxies in the full NSA-VLASS-WISE sample. This result matches with the $g-r$ color analysis that shows that the dwarf galaxies are bluer in color than the full sample, indicative of more recent SF.
We note, of course, that several galaxies in of the AGN population have upper limit detections in WISE, meaning their true location in the mass-SFR plot may be lower than what is shown.

Overall, we find that the dwarf galaxies hosting AGNs do not follow the general trends of the full population of AGN hosts. The radio sources are in general less luminous and the dwarf galaxy hosts are bluer, less centrally concentrated, and more abundant in SF than the more massive AGN hosts. 

\subsection{Comparison to Other AGN Detection Methods}
\label{subsec:dwarfs_other_agn_methods}
\begin{figure}
 \includegraphics[width=\columnwidth]{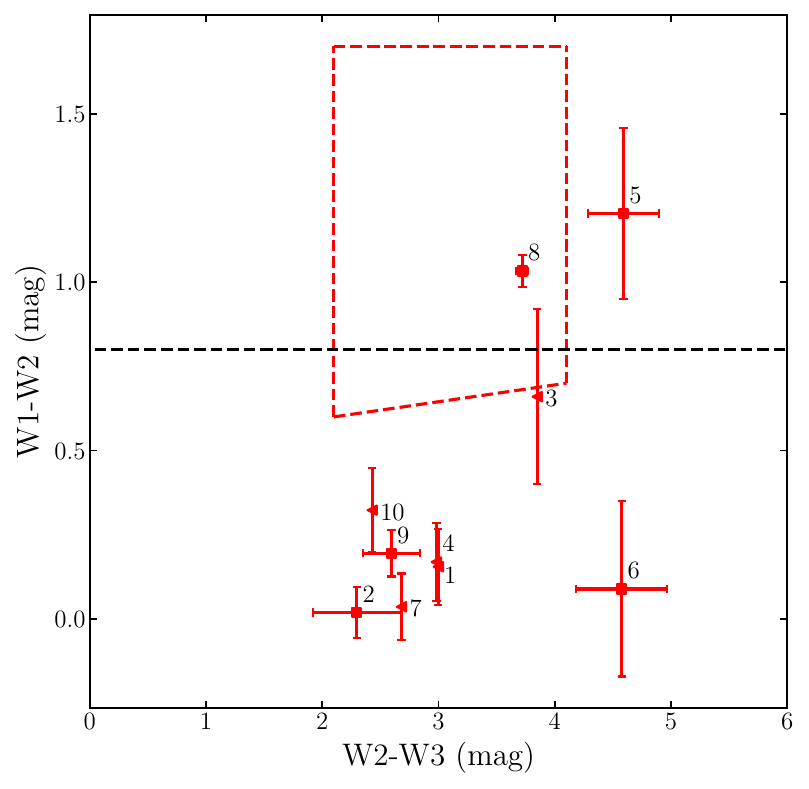}
 \caption{The WISE color-color diagram for the 10 radio-excess AGNs in dwarf galaxies. The ID number associated with each point is shown. The same AGN cutoff and selection boxes from Figures \ref{fig:wise_color} and \ref{fig:reas_wise_color} are also shown. {The error bars for the WISE colors are shown, and galaxies with upper limits in W3 are shown as triangles.} Only 2 of the dwarf galaxies in our AGN sample have mid-IR colors indicative of AGNs: {ID 8 is found within the \cite{jarrett2011} selection box and above the \cite{stern2012} cutoff and ID 5 is only found above the \cite{stern2012} cutoff. {ID 3 could also be found within the \cite{jarrett2011} selection box, due to its large uncertainty in its WISE colors.}}  
 } 
\label{fig:dwarf_wise_color_color}
\end{figure}

We analyze the radio-excess AGNs in dwarf galaxies in the WISE color-color space to identify which galaxies are classified as mid-IR AGN hosts (see Figure \ref{fig:dwarf_wise_color_color}). We do not restrict by WISE S/N, resulting in several galaxies with S/N $<$ 5 in the W3 band. We plot these points with low S/N with their upper limit magnitudes in Figure \ref{fig:dwarf_wise_color_color}.  

We find that the radio-excess AGNs in dwarf galaxies are not consistently identified as AGNs by the mid-IR selection methods: only 2 of the 10 AGNs are found within the \cite{jarrett2011} box or above the \cite{stern2012} cutoff. 
These results match well with the full population (see Figure \ref{fig:reas_wise_color}) which found a similar percentage of radio-excess AGNs to be identified as AGN with mid-IR observations.

\begin{figure}
 \includegraphics[width=\columnwidth]{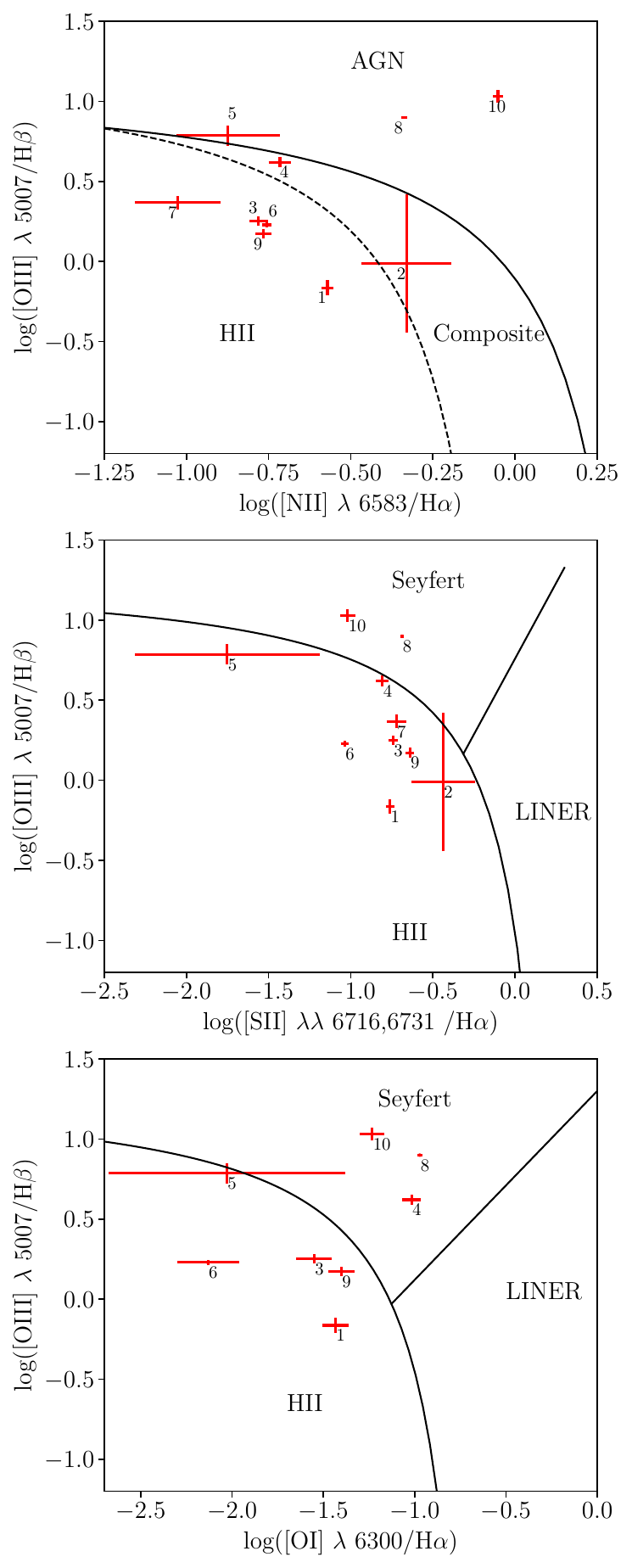}
 \caption{Optical emission line diagnostic diagrams for the  dwarf galaxies with radio-excess AGNs. ID numbers for each of the galaxies in our sample are shown, along with the error bars in the measurements. The narrow line emission measurements come from the SDSS spectral fits and the divisions between regions of the diagram come from the classification scheme outlined in \cite{kewley2006}. Most of the AGN hosts lie in the SF sections of the diagrams.}
\label{fig:bpt}
\end{figure}

We also analyze the 10 radio-excess AGNs in dwarf galaxies with narrow line diagnostic diagrams, using the fluxes taken from the SDSS spectral fit (Figure \ref{fig:bpt}). All 10 galaxies appear on the [O III]/H$\beta$ versus [N II]/H$\alpha$  diagram and the [O III]/H$\beta$ versus [S II]/H$\alpha$ diagram, but only 8 appear on the [O III]/H$\beta$ versus [O I]/H$\alpha$ diagram since IDs 2 and 7 do not have reliable OI emission line fits. 
Most of the galaxies lie in the HII region of the emission lines diagrams. These apparently contradictory results align well with the results from \cite{reines2020}, which also found a majority of radio-detected AGNs in dwarf galaxies to not lie in the AGN region of the BPT diagrams. 

Overall, the comparison between the AGN selection methods show that the radio-excess method for finding AGNs does not correlate well with mid-IR and optical methods. Many AGNs are only identified as AGNs in the radio regime, and only 2 (IDs 5 and 8) are identified in AGNs in the radio, optical, and IR regimes. 



\subsection{Notes on Individual Galaxies}
\label{subsec:dwarfs_individual}
Given the small number of dwarf galaxies hosting radio-excess AGNs, we consider the properties and previous research preformed on each galaxy individually, as follows:

\textit{ID 1 (J090313.1+482415.1):} ID 1 has been previously identified by \cite{reines2020} as a potential host of a wandering BH {(see ID 25 in their paper).} The VLASS detection is located 2\farcs1 away from the coordinates of its associated galaxy in SDSS and the radio source appears by eye to be distinctly offset from the center of the galaxy (see Figure \ref{fig:decals_pics}). ID 1 was observed with the Very Large Baseline Array (VLBA) by \cite{sargent2022}, but no detections were found. However, \cite{sargent2022} identified the radio source as an AGN by comparing the expected SFR from WISE and GALEX data to the expected SFR from FIRST data. The radio-predicted SFR was found to be excessively high, indicating that the radio emission was not likely to have come from SF alone. Therefore, this paper’s classification of ID 1 as a radio-excess AGN is consistent with the classifications from \cite{reines2020} and \cite{sargent2022}.
\cite{eftekhari2020} also studied ID 1 and proposed that it could be a persistent radio source associated with a fast radio burst (FRB).

\textit{ID 2 (J092157.8+330426.0):} Of the 12 dwarf galaxies with radio-excess AGNs, ID 2 has the largest offset between the radio source and the optical center of its host galaxy, with an offset of 2\farcs3. 
Like ID 1, the DECaLS images (Figure \ref{fig:decals_pics}) reveal that the radio source is not located in the nucleus of its host galaxy. The $grz$-band image from DECaLS shows a point-like object that is prominent in the $z$ band at the same location as the radio source. Given its distance from the galactic nucleus, the source is likely to be either a background AGN or a wandering BH in the dwarf galaxy. 
The [NII]/H$\alpha$ narrow line diagnostic diagram classifies the dwarf galaxy as a composite. 
{The radio source is undetected in FIRST, which can be explained due to the difference in sensitivities between FIRST and VLASS. VLASS reports the radio source to have a peak flux of 0.82 mJy/beam, but the catalog detection limit for FIRST in this region of the sky was 0.89 mJy/beam. {The radio source was also not detected in the first epoch of VLASS, but was in the second epoch, when the depth of the survey increased.}} 




\textit{ID 3 (J093138.5+563318.7):} ID 3 was studied by \cite{reines2020} and was identified as a radio AGN (see ID 33 in their paper). \cite{sargent2022} also conducted a follow-up study on ID 3: while they found no VLBA detections, ID 3 was classified as an AGN based on the same SFR comparisons performed on ID 1. The classification of ID 3 in this paper is consistent with \cite{reines2020} and \cite{sargent2022}. Although the angular offset is small (1\farcs4), visual inspection of the radio source finds it to be located at the optical outskirts of its host galaxy, making it a likely candidate for a wandering BH. 
\cite{eftekhari2020} also studied ID 3 and found its radio emission to be consistent with {a persistent radio source associated with} an FRB.

\textit{ID 4 (J101747.1+393207.8):} VLASS found ID 4 to have a peak flux of 1.18 mJy/beam. Even though ID 4 lies in the FIRST footprint in a region of sky with a FIRST detection limit of 1.03 mJy/beam, there were no associated FIRST detections. {ID 4 was also not detected in the first epoch of VLASS.} 
The optical spectrum of ID 4 was studied in \cite{reines2013} and was classified as a composite galaxy using BPT diagrams (ID 66 in their paper). Similarly, in the narrow-line emission diagrams made with SDSS spectral fits (see Figure \ref{fig:bpt}), we find ID 4 to be classified as a composite galaxy in the [NII]/H$\alpha$ diagram. However, we also find it to be classified as a Seyfert in the [OI]/H$\alpha$ diagram, and as an HII region by the [SII]/H$\alpha$ diagram. Though the optical classifications may be unclear, the radio classification clarifies that this object is most likely a variable AGN. 
The morphology of ID 4 was also studied by \cite{kimbrell2021} (RGG 66) using HST observations, and was found to be an irregular galaxy with signs of undergoing a merger. Follow-up {\it Chandra} observations reveal a luminous X-ray AGN (Kimbrell \& Reines, in preparation).

\textit{ID 5 (J102038.7+212806.4):} ID 5 is classified as an AGN by the [NII]/H$\alpha$ narrow line diagnostic diagram (Figure \ref{fig:bpt}) and has WISE colors consistent with an AGN (Figure \ref{fig:dwarf_wise_color_color}). 
ID 5 was not included in \cite{reines2020} because its redshift (z $=0.13$) is larger than any of the redshifts included in the version of the NSA utilized by \cite{reines2020}. 
ID 5 was identified as a Type 2 AGN with a potential double-peaked [OII] spectrum by \cite{smith2010}.


\textit{ID 6 (J113910.9+110119.6)}: 
ID 6 was not included in the \cite{reines2020} because it had a large enough redshift (z$=$0.08) to not be included in the version of the NSA utilized by \cite{reines2020}. The radio source appears to not be centered on its host galaxy, meaning that it could be a wandering BH.
{The radio source is extended, and from the VLASS image appears to have radio jets (Figure \ref{fig:vlass_pics}).}

\textit{ID 7 (J130616.2+360847.4):} ID 7 is the reddest of the galaxies in our sample of 13, with a $g-r$ color of 0.99. The radio source is centered on its host galaxy, making it a nuclear BH.

\textit{ID 8 (J133245.6+263449.3)}: \cite{reines2013} previously identified ID 8 using optical selection methods (ID 24 in their paper). It was considered as a target for high resolution VLA observations in \cite{reines2020} but was cut due to scheduling priorities. ID 8 was also observed by the Chandra X-ray Observatory in \cite{latimer2021b} and its X-ray emission was higher than expected from X-ray binaries (XRBs), indicating that it was likely created by an AGN (ID 5 in their paper). Furthermore, ID 8 is classified as an AGN by WISE colors (Figure \ref{fig:dwarf_wise_color_color}) and by the narrow line emission diagnostics (Figure \ref{fig:bpt}). Its classification as an AGN in the radio regime matches well with the classification in the IR, optical, and X-ray regimes.

\textit{ID 9 (J162244.6+321259.9):} ID 9 was considered as a target for high resolution VLA observations in \cite{reines2020}, but was cut due to scheduling priorities. 
ID 9 was studied by \cite{ofek2017}, where it was shown to be consistent with a persistent radio source associated with an FRB. {Although it was detected in FIRST and the second epoch of VLASS, ID 9 was not detected in the first epoch of VLASS. 
}

\textit{ID 10 (J232135.2+121306.9):} ID 10 is classified as an AGN by all 3 narrow line diagnostic diagrams. ID 10 has a peak VLASS flux of 2.12 mJy/beam, but did not have corresponding FIRST observations, despite being in a region in the sky with a FIRST detection limit of 0.97 mJy/beam. {ID 10 also does not have any corresponding detections in the first epoch of VLASS.} ID 10 is therefore a variable radio source. 

\subsection{Comparison to \cite{reines2020}}
\label{subsec:reines2020}

Although this paper employed methods similar to those in \cite{reines2020}, the samples of radio AGNs in dwarf galaxies discovered between the two papers differ substantially. Of the 10 dwarf galaxies with radio-excess AGNs in our final sample, only two (IDs 1 and 3) were also identified as radio AGNs by \cite{reines2020} (IDs 25 and 33 in their paper). 
Most of the AGNs in our sample but not in \cite{reines2020} were found due to a difference in data sets (see \S\ref{sec:Data}). 
Three of the galaxies here (IDs 5, 6, and 7) were not included in the older version of the NSA utilized by \cite{reines2020}. Three other galaxies (IDs 2, 4, and 10) were not detected in FIRST, which was a requirement of \cite{reines2020};
ID 2 has a VLASS flux density below the detection threshold for FIRST, and the other galaxies are likely to be variable sources that also fell below the detection limit of FIRST. The remaining 2 galaxies (IDs 8 and 9) were not included in \cite{reines2020} due to VLA scheduling priorities. They both had FIRST detections, but were ultimately not followed-up with high resolution VLA observations. {ID 9 notably does not have any detections in epoch 1 of VLASS, but does appear in the epoch 2 catalog.} 

We also investigate why many of the AGNs found in \cite{reines2020} were not also discovered in this paper. With this paper using a more inclusive version of the NSA than \cite{reines2020} and with VLASS having better resolution and sensitivity that covered all of the sky studied by FIRST, it is interesting to find that of the 13 radio AGNs in bona fide galaxies reported in \cite{reines2020}, only 2 were considered in our final dwarf galaxy sample. Here, we investigate why the other 11 were not included in our sample.

One galaxy (ID 64 in \citealt{reines2020}) was discovered by our method, but was eliminated from our final sample since follow-up spectroscopy revealed it to be a background source (see \S\ref{subsec:select_dwarfs_agns}).
Another galaxy (ID 92 in \citealt{reines2020}) was identified as an AGN by \cite{reines2020} but was later identified as being consistent with SF by \cite{sargent2022}. We also found this galaxy to be consistent with SF in this paper. 
Four of the galaxies had optical-radio offsets above 2\farcs5 in \cite{reines2020} (IDs 2, 28, 48, 65 in their paper), and so were not found with our cross-matching radius of 2\farcs5. We choose not to increase our cross-matching radius to include these sources however, since it would increase the probability of including background sources. Indeed, \cite{sargent2022} detected these 4 AGNs with VLBA observations and argued that, due to their large offset, they were statistically likely to be background sources.

The remaining galaxies not in this paper (IDs 6, 26, 77, 82, 83 in \citealt{reines2020}) were not included because they did not have any corresponding detections in the VLASS source catalog, 
{even though they were within the VLASS footprint}. Two of the sources (IDs 6 and 82) have 1.4 GHz flux densities from FIRST around 1 mJy (1.47 and 0.99 mJy, respectively), and the other sources have flux densities from FIRST above 4 mJy \citep{reines2020}. 
We compare these flux densities to the detection limits of VLASS.
To be listed in the VLASS catalog, a source has to have a peak flux density 5 times greater than the rms noise of the image. Given that VLASS has median rms noise levels of 128$-$145 $\mu$Jy beam$^{-1}$ \citep{gordon2021}, the minimum flux for a detection to be included in the VLASS catalog is in the range of 640$-$725 $\mu$Jy beam$^{-1}$. 
The fact that the sources are not detected in VLASS indicates that the flux of the sources has changed drastically between the time of the FIRST and VLASS observations, especially for the sources with FIRST flux densities above 4 mJy.  Therefore, they are likely to be variable sources. {We explore the nature of the variable sources in the next section.}  

\subsection{Variable Sources}
\label{subsec:variable}
In our individual analysis of the dwarf galaxies in our sample, we identified 2 galaxies (IDs 4 and 10) that were within the FIRST footprint and had large enough radio fluxes in VLASS to be detected in FIRST, but had no corresponding FIRST detections. In comparing our dwarf galaxy sample to that from \cite{reines2020}, we found the opposite: there were 5 galaxies (IDs 6, 26, 77, 82, and 83 from \citealt{reines2020}) that had large enough fluxes in FIRST to be detected in VLASS, but had no corresponding observations in VLASS, despite being in the footprint of VLASS. {We also identified one galaxy (ID 9) that had a FIRST detection and VLASS detection in epoch 2, but did not have a VLASS detection in epoch 1.}  We consider these sources to be radio variables and we examine possible causes of their variability.

Two possible sources of radio variability are SNe and gamma-ray bursts (GRBs). While SNe were shown to be unlikely causes of the radio emission in our sample in \S\ref{subsec:select_dwarfs_agns}, radio-loud supernovae can still create luminosities comparable to those exhibited by our variable sources. PTF11qcj, the radio-loud supernova we eliminated from our sample (\S\ref{subsec:select_dwarfs_agns}), has a 3 GHz luminosity of $\sim$10$^{22}$ W Hz$^{-1}$ in VLASS, which is a similar magnitude to the radio luminosities of the the variable sources in our dwarf galaxy sample (see Table \ref{tab:dwarf_agns}). The variable sources from \cite{reines2020} also had similar radio luminosities to PTF11qcj, with 1.4 GHz luminosities of 10$^{20.9}-10^{22.3}$ W Hz$^{-1}$ from FIRST. 

The radio luminosities of our variable sources could also be consistent with GRBs. On-axis GRBs have peak luminosities of $\sim$10$^{24}$ W Hz$^{-1}$ \citep{chandra2012}, which is a magnitude larger than the luminosity of any of the variable sources in this paper or \cite{reines2020}. {However, the radio emission from GRBs is variable over a timescale of around 1–2 weeks \citep{pietka2015}. Since FIRST, VLASS epoch 1, and VLASS epoch 2 observations were taken years apart from each other, if a source is detected in more than one of these surveys, it is unlikely to be a GRB (e.g., ID 9 here and IDs 6 and 26\footnote{Note that ID 26 in \citealt{reines2020} is an optically-selected AGN (ID 9 in \citealt{reines2013}).} from \citealt{reines2020}). 
}

Tidal disruption events (TDEs; e.g., \citealt{komossa2015}, and references therein) are another possible source of radio variability, as stars that are ripped apart and accreted onto SMBHs create radio emission.
The most luminous radio TDE, Swift J1644+57, peaked at a radio luminosity of $\sim$$10^{25}$ W Hz$^{-1}$ \citep{eftekhari2018}, which is larger than any of the radio luminosities of the variable sources in this paper or \cite{reines2020}. TDEs are less likely to occur with large SMBHs due their weaker tidal fields \citep{hills1975}, but the smaller BHs we have discovered in dwarf galaxies could be likely candidates for TDEs. However, while TDEs are a possible source of the variability, the current literature agrees that TDEs with powerful relativistic jets are rare (e.g., \citealt{vanvelzen2018}), and constitute a small percentage of the known TDE population (\citealt{alexander2020}, and references therein).

While SNe, GRBs, and TDEs could be the sources of radio variability, AGNs are more likely candidates, since the literature has shown that the radio variable population is dominated by AGNs (e.g., \citealt{carilli2003}; \citealt{bannister2011}; \citealt{thyagarajan2011}; \citealt{frail2012}; \citealt{bell2015}; \citealt{mooley2016}).  The radio variability of AGNs can be caused by mechanisms intrinsic to the AGN or by extrinsic propagation effects, such as interstellar scattering (\citealt{bignall2015} and references therein). Follow-up observations at different radio frequencies may be performed on the variable objects noted here to create multi-band spectral energy distributions (SEDs) that will help determine if the variability of these objects is consistent with transient phenomena, extrinsic propagation effects, or AGNs (as was done in \citealt{nyland2020}).


\section{Conclusions and Discussion} 
\label{sec:Conclusions}
We present a large-scale search for radio-excess AGNs in dwarf galaxies using data from the second epoch of VLASS. We first develop a criterion to distinguish between SF-related radio emission and AGN radio emission by analyzing the IRRC for $\sim 7,000$ galaxies in the NSA ($z \leq 0.15$) across the mass scale with VLASS and WISE detections (the NSA-VLASS-WISE sample). Using our derived $2\sigma$ threshold of $q < 1.94$, we find 723 galaxies with radio-excess AGNs within this NSA-VLASS-WISE sample. We have compiled catalogs of {6,904 radio-excess AGNs (including galaxies with low WISE S/N) and 6,049 SF-consistent galaxies} in VLASS and make these available to the community (see Appendix).

Applying the criterion to bona fide dwarf galaxies, we find 10 that are apparent hosts of radio-excess AGNs. Our statistical analysis suggests that the majority of the radio-excess AGNs in our sample are indeed associated with the dwarf host galaxies, but we cannot completely rule out partial contamination from background AGNs. Our main results on the dwarf galaxy sample are summarized below:


\begin{enumerate}

\item The dwarf galaxies with radio-excess AGNs starkly differ from the majority of AGN hosts in the NSA-VLASS-WISE sample. 
The dwarf galaxy hosts have bluer colors and larger amounts of star formation than typical more massive radio-excess AGN hosts. Additionally, the dwarf galaxy hosts are not as centrally concentrated as the more massive hosts. This suggests that the dwarf hosts of radio-excess AGNs are similar to the general population of dwarf galaxies (e.g., Figure \ref{fig:dwarf_mass_gr}).  
\item Five of the ten dwarf galaxies have evidence for hosting AGNs based on optical emission line diagnostic diagrams. Two of these also have mid-IR colors indicating the presence of an AGN. Importantly, the emission lines confirm the AGNs are associated with the dwarf galaxies and not background sources.




\item {When we compare our sample of AGNs in dwarf galaxies to the sample from \cite{reines2020}, we find evidence for {8} galaxies with variable radio sources. Follow-up observations will be critical for discerning the cause of the variability of these sources.}

\item Five of the dwarf galaxies with radio-excess AGNs are uniquely identified as AGNs in this paper
and are prime candidates for follow-up multi-wavelength observations. 
\end{enumerate}

The radio-excess AGNs in the dwarf galaxies presented here are probable candidates for being IMBHs. If we assume the mass of the AGNs to follow the BH mass to total stellar mass ($M_{BH}-M_{*}$) relation from \cite{reines2015}, 
we expect the BHs in these galaxies to have masses of $M\sim10^{4.9}-10^{5.8}\;M_\odot$, with a median value of $10^{5.7}\;M_\odot$. We stress that there is a large amount of uncertainty in these estimates: the scatter in the \cite{reines2015} is quite large (0.55 dex).



This work shows the potential for finding new AGNs in dwarf galaxies using large-scale radio surveys. 
As such, upcoming radio facilities, such as the next generation VLA, 
could be utilized to efficiently find large numbers of radio-excess AGNs in dwarf galaxies (see \citealt{plotkin2018}). Such observations will better constrain the occupation fraction of IMBHs in dwarf galaxies, which will consequently constrain the seeding models for SMBHs.


\section*{Acknowledgments}
We thank the anonymous reviewer whose suggestions helped improve and clarify the manuscript. A.E.R. gratefully acknowledges support for this work provided by NSF through CAREER award 2235277 and NASA through EPSCoR grant No. 80NSSC20M0231.

\section*{Appendix}
We provide the catalogs of all the radio-excess AGNs and the SF-consistent galaxies in the NSA and VLASS as machine-readable tables, and show the first 5 rows of each catalog in Tables A.1 and A.2. {The AGN catalog includes galaxies that have WISE observations with low S/N. These galaxies have upper limits for their IR luminosities, but even if their IR luminosities (and the corresponding $q$ values) are lower than the reported values, their emission is still consistent with radio-excess AGNs.} 

{We show the spectra of the 10 dwarf galaxies that we identified as strong candidates for hosting radio-excess AGNs in Figure A.1.}
\newpage
\begin{deluxetable}{ccccccccccccc}
\rotate
\tablenum{A.1}
\tablecaption{Radio-excess AGNs in the NSA and VLASS}
\tablewidth{0pt}
\tablehead{
\colhead{Name} & \colhead{NSA ID} & \colhead{R.A.} &\colhead{Dec.} & \colhead{$z$} & \colhead{log(M$_{*}$)} & \colhead{$g-r$} &  \colhead{Offset} & \colhead{S$_{3\;\mathrm{GHz}}$} & \colhead{log(L$_{3\;\mathrm{GHz}}$)} & \colhead{$\Psi_{maj}$} & \colhead{log(L$_{1.4\;\mathrm{GHz}}$)} & \colhead{log(L$_{\mathrm{TIR}}$)}}
\colnumbers
\startdata
    J00006.99+081645.0 &  613637 &  0.02933 &   8.27922 &  0.0394 &  11.34 &  0.81 & 0.28 &   50.90(0.41) &  23.19(0.00) &  0.6(0.0) & 23.35(0.00) & $<$36.08\\
  J000010.45+073246.9 &  590457 &  0.04357 &   7.54638 &  0.1365 &  11.14 &  0.78 &  0.40 &    5.97(0.64) &  23.29(0.05) &  3.7(0.4) & 23.28(0.05) & $<$37.05\\
  J000022.36+203412.4 &  605059 &  0.09318 &  20.57011 &  0.0734 &  10.84 &  0.80 &  0.55 &    2.01(0.45) &  22.33(0.10) &  4.6(1.0) & 22.56(0.10) & $<$36.60\\
  J000059.28-020941.3 &  586592 &  0.24700 &  -2.16151 &  0.1154 &  11.01 &  0.78 &  0.33 &    9.50(0.29) &  23.44(0.01) &  0(0.1) & 23.99(0.01) & $<$37.20\\
  J000114.33+061422.0 &  588788 &  0.30976 &   6.23946 &  0.0963 &  10.95 &  0.78 &  0.38 &   21.66(0.35) &  23.56(0.01) &  1.4(0.0) & 23.48(0.01) & $<$36.78\\
\enddata
\tablecomments{Column 1: galaxy name. Column 2: NSA ID (version v1\textunderscore0\textunderscore1). Column 3: R.A. of galaxy from the NSA, in units of degrees. Column 4: dec. of galaxy from the NSA, in units of degrees. Column 5: redshift given by NSA. Column 6: log galaxy mass given by the NSA, in units of $M_\odot$, Column 7: $g - r$ color given by the NSA. Column 8: Offset between the galaxy coordinates in the NSA and the radio source location in VLASS, in units of arcsec. Column 9: integrated radio flux density of radio source from VLASS, in units of mJy at 3 GHz.  
Column 10: log 3 GHz luminosity of radio source from VLASS, in units of W Hz$^{-1}$. Column 11: deconvolved major axis size of radio source from VLASS, in units of arcsec. Column 12: log 1.4 GHz luminosity of radio source from FIRST, in units of W Hz$^{-1}$. Galaxies without FIRST detections have their radio luminosities estimated using a spectral index of $\alpha=-0.7$. Column 13: TIR luminosity, estimated using W4 magnitudes from WISE, in units of W. Upper limits are indicated. Errors for the last 5 columns are given in parentheses.    }
\end{deluxetable}
\label{tab:agns_all}
\begin{deluxetable}{ccccccccccccc}
\rotate
\tablenum{A.2}
\tablecaption{SF-consistent Galaxies in the NSA and VLASS}
\tablewidth{0pt} 
\tablehead{
\colhead{Name} & \colhead{NSA ID} & \colhead{R.A.} &\colhead{Dec.} & \colhead{$z$} & \colhead{log(M$_{*}$)} & \colhead{$g-r$} &  \colhead{Offset} & \colhead{S$_{3\;\mathrm{GHz}}$} & \colhead{log(L$_{3\;\mathrm{GHz}}$)} & \colhead{$\Psi_{maj}$} & \colhead{log(L$_{1.4\;\mathrm{GHz}}$)} & \colhead{log(L$_{\mathrm{TIR}}$)}}
\startdata
  J000002.29-042805.0 &  639204 &  0.00993 &  -4.46780 &  0.1028 &  10.89 &  0.58 & 0.13 &  4.60(0.58) &  22.96(0.06) &  2.2(0.5) & 23.03(0.06) & 38.11(0.03)\\
  J000009.19+252101.0 &  639211 &  0.03645 &  25.34987 &  0.0594 &  10.79 &  0.66 & 0.22  &  2.89(0.46) &  22.31(0.07) &  3.2(0.5) & 22.54(0.07) & 37.78(0.02)\\
  J000034.96+173405.9 &  608760 &  0.14570 &  17.56832 &  0.0996 &  10.79 &  0.64 &  0.08  & 1.34(0.27) &  22.42(0.09) &  2.6(0.6) & 22.65(0.09) & 37.60(0.04)\\
  J000040.26-054100.9 &  639226 &  0.16772 &  -5.68361 &  0.0944 &  10.34 &  0.47 &   0.37  &2.36(0.33) &  22.61(0.06) &  3.6(0.7) & 22.78(0.06) & 38.36(0.01)\\
  J000054.50+183021.9 &  639233 &  0.22617 &  18.50613 &  0.0563 &   9.68 & -0.01 &    0.65 &2.08(0.28) &  22.12(0.06) &  2.4(0.4) & 22.35(0.06) & 38.24(0.01)\\
\enddata
\tablecomments{Same columns and units as Table A1. Tables A.1 and A.2 are available in their entirety in machine-readable form.}
\end{deluxetable}
\label{tab:sf_all}
\clearpage

\renewcommand{\thefigure}{A.1}
\begin{figure}
\includegraphics[width=\textwidth]{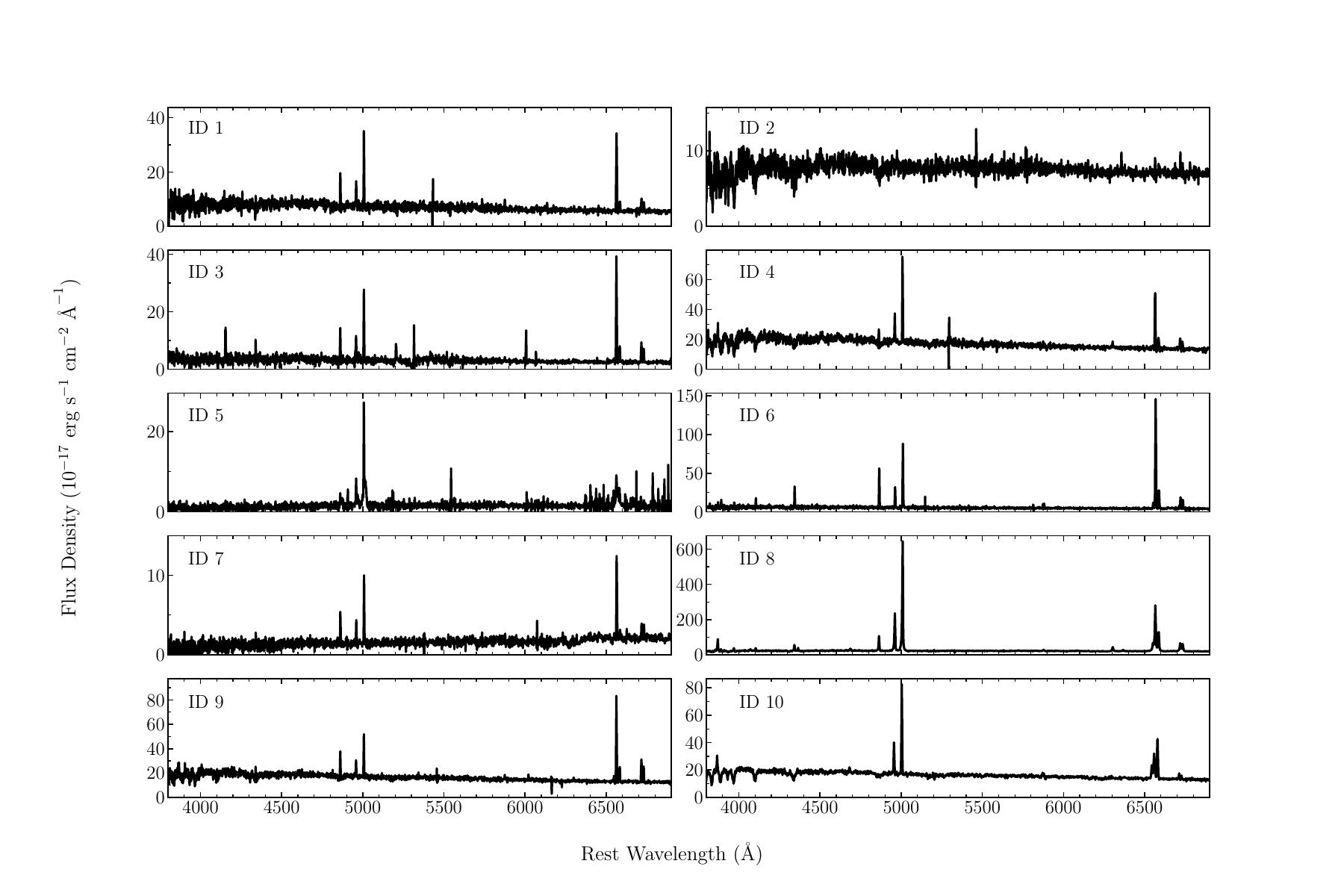} 
\label{fig:spec_10agn}
\parbox{\textwidth}{\caption{{SDSS redshift-corrected spectra of the 10 dwarf galaxies that are strong candidates for hosting radio-excess AGNs.}}}
\end{figure}


\bibliography{agn_research_bib}{}
\bibliographystyle{aasjournal}
\nocite{allwisecat}

\end{document}